\documentclass[aps, prd, twocolumn, showpacs, superscriptaddress, groupedaddress]{revtex4-1}

\usepackage{graphicx}	
\usepackage{multirow}
\usepackage{amssymb}
\usepackage{dcolumn}
\usepackage{color}
\usepackage{subfigure, rotating, bm, array}
\usepackage{comment}
\usepackage[pagebackref=false, colorlinks=true]{hyperref}
\hypersetup{
linkcolor=blue,     
citecolor=blue,     
urlcolor=blue}      
\begin{document}
\title{Comparing thin accretion disk properties of naked singularities and black holes}
\author{Divya Tahelyani}
\email{tahelyanidivya118@gmail.com}
\affiliation{International Center for Cosmology, Charusat University, Anand, GUJ 388421, India}
\author{Ashok B. Joshi}
\email{gen.rel.joshi@gmail.com}
\affiliation{International Center for Cosmology, Charusat University, Anand, GUJ 388421,  India}
\author{Dipanjan Dey}
\email{deydipanjan7@gmail.com}
\affiliation{International Center for Cosmology, Charusat University, Anand, GUJ 388421, India}
\author{Pankaj S. Joshi}
\email{psjcosmos@gmail.com}
\affiliation{Cosmology Centre, Ahmedabad University, Ahmedabad, GUJ 380009, India\\
International Center for Cosmology, Charusat University, Anand, GUJ 388421,  India}
\date{\today}
\begin{abstract}
In the present paper, we study the thermal properties of the geometrically thin accretion disks surrounding the null naked singularity (NNS) and the first type of Joshi-Malafarina-Narayan (JMN1) spacetimes and compare the results with the accretion disk around equally massive Schwarzschild black hole. First, we examine the properties of the circular orbits in these spacetimes. The emitted flux, radiation spectra, disk efficiency, and temperature distribution on the disk surface are then investigated. The efficiency of the conversion of the accreting matter into radiation is found to be substantially higher for naked singularities than that for black holes. We also verify that the flux radiated from the disk surface is greater for null and JMN1 naked singularities than black holes. Hence, the accretion disks around naked singularities are much more luminous than the black holes of the same mass and accretion rate. In the luminosity spectra of the NNS, we find that the significant contribution of the low-frequency is coming from the nearby regions of the NNS. Furthermore, the spectral luminosity distribution for the ``non-zero torque" at the inner boundary is also analysed by the inclusion of the non-zero torque value at the inner edge of the disk. The slopes of the luminosity distribution with respect to frequency for naked singularity spacetimes differ significantly from those of black holes. These unique features of the naked singularities serve as an effective tool to distinguish them from the equally massive black holes.

\bigskip
Key words: Black hole, Naked singularity, Accretion disk.
\end{abstract}

\maketitle

\section{introduction}
The resources for understanding astrophysical phenomena have increased significantly in the wake of recent experimental breakthroughs in astrophysical observations, such as gravitational wave detection by the LIGO \& Virgo Collaboration~\cite{LIGOScientific:2016aoc} and observation of the black hole image in Galaxy M87~\cite{EventHorizonTelescope:2019dse, EventHorizonTelescope:2019uob, EventHorizonTelescope:2019jan, EventHorizonTelescope:2019ths, EventHorizonTelescope:2019pgp, EventHorizonTelescope:2019ggy} by the Event Horizon Telescope group. Meanwhile, the next generation Event Horizon Telescope (ngEHT) group is also attempting to capture the sharpest image of the galactic center of our Milky Way. It is widely believed that the compact object at the center of almost every galaxy is a supermassive black hole. Nevertheless, there are numerous astrophysical compact objects that can replicate the observational properties of black holes, such as naked singularities, gravastars, wormholes, and so on. In~\cite{Vagnozzi:2019apd, Shaikh:2019hbm, Gyulchev:2019tvk, Shaikh:2019fpu, Dey:2013yga, Dey+15, Shaikh:2018lcc, Joshi2020, Paul2020, Dey:2020haf, Dey:2020bgo, bambi_2013a, ohgami_2015}, authors have shown that apart from the black holes, different horizon-less compact objects can also produce shadows. 

In general relativity, the dynamical equations describing gravitational collapse do not require that the eventual end state of the collapse of a massive matter cloud must be a black hole. Depending upon the initial conditions with which the collapse is initiated, the gravitational collapse of a massive star may ultimately result in a black hole or a naked singularity. As we know, the cosmic censorship conjecture rules out the possibility of horizon-less compact objects~\cite{penrose}. However, several pieces of literature propose that the continuous gravitational collapse of an inhomogeneous matter cloud can result in a naked singularity~\cite{eardley, christodoulou, joshi, goswami, joshi2, vaz, jhingan0, deshingkar, mena, magli1, magli2, giambo1, harada1, harada2, joshi7, mosani1, mosani2, mosani3, mosani4, mosani5}. The theoretical investigations alone are not sufficient to confirm whether the naked singularities exist or not in nature. Therefore, studying their distinct observational properties is necessary to distinguish them from black holes. Moreover, the causal structure of the singularity also remains a mystery. As per general relativity, spacetime singularities can be classified into three categories based on their causal structure: spacelike, nulllike, and timelike. A spacelike singularity is causally detached from other spacetime points, while nulllike and timelike singularities are causally related to the other points of spacetime. The difference in the spacetime geometries around black holes and naked singularities may produce differences in their observable signatures, allowing us to discriminate between these two classes of compact objects. The causal nature of the compact object can also be probed by the orbital dynamics of the stars orbiting that compact object. There are several papers on the detailed study of the orbital precession of timelike geodesics around various compact objects~\cite{Dey:2020haf, Bambhaniya:2019pbr, Dey:2019fpv}, where it is demonstrated that the orbital motion around a naked singularity can be substantially different from the orbital motion around a Schwarzschild black hole of the same mass.

The availability of an abundance of data in the electromagnetic realm has piqued the interest of researchers all around the world because such a study might help us better comprehend the nature of compact objects at the centers of active galactic nuclei (AGN) or in the X-ray binaries. In this paper, we attempt to distinguish naked singularities from black holes by investigating the electromagnetic properties of the thin accretion disk around them. The majority of the astrophysical structures grow via accretion. The particles orbiting an astrophysical body lose angular momentum and energy as they flow in, which prevents them from directly falling into the central body and settling into a disk-like structure. The first explicit non-relativistic study of the accretion disks around compact objects was given by Shakura and Sunyaev~\cite{Shakura:1973}. Later, Novikov and Thorne~\cite{Novikov}, and Page and Thorne~\cite{Page:1974he} investigated the general relativistic model of the accretion disk. There is much literature in which the electromagnetic properties of the accretion disks have been studied for different compact objects~\cite{Liu:2020vkh, Liu:2021yev, Joshi:2013dva, Bambhaniya:2021ugr, Rahaman:2021kge, Harko:2008vy, Harko:2009xf, Kovacs:2010xm, Harko:2009gc}. In~\cite{Guo:2020tgv, Chowdhury:2011aa}, the authors have demonstrated that the thermal properties of the accretion disk around a naked singularity can differ considerably from those of a black hole of the same mass.

The matter density around the central supermassive object in the galaxy is very high. Hence, describing its surrounding geometry using vacuum solutions of Einstein's field equations may not always yield a good approximation. To account for the effect of the surrounding matter on the galactic geometry, here we consider two non-vacuum solutions, namely, NNS and JMN1 spacetimes. These spacetimes contain the matter distribution around the central singularity. We carry out a comparative study of the accretion properties of the thin accretion disk surrounding a Schwarzschild black hole, an NNS~\cite{Joshi:2020tlq}, and a JMN1 naked singularity~\cite{Joshi:2011zm}. The Schwarzschild and NNS spacetime have, respectively, a spacelike and nulllike singularity at their center. On the other hand, JMN1 spacetime possesses a timelike or nulllike singularity at its centre, depending on the parameter space. Hence, we particularly try to analyse the different astrophysical signatures provided by the radiation emitted by accretion disks around spacelike, nulllike, and timelike singularities. We study the radiated energy flux, spectral luminosity distribution, accretion efficiency, and the temperature distribution on the disk surface for equally massive black holes and naked singularities. From the analysis of the radiating flux and spectral luminosity, we show that the accretion disks surrounding naked singularities are more luminous than that of the black holes.  In the Novikov-Thorne accretion disk model, several reasonable assumptions are made. As one of the assumptions, in general, a ``zero-torque" condition is considered at the inner boundary of the accretion disk to compute the time-averaged flux~\cite{Novikov, Page:1974he}, assuming that there is a minimal quantity of material inside the innermost circular orbit (ISCO) that can ``torque up" the material immediately outside ISCO. This nearly vanishing torque could be a reasonable assumption for spacetimes possessing ISCO. However, if spacetime has circular orbits all the way up to the zero radial distance, there is a possibility of the existence of non-zero torque owing to the viscosity between the fluid layers near the inner edge induced by the strong gravity near the singularity. This torque might be minimal, but it may have some non-zero value. To address this possibility, we also examine the differences observed in the spectra of the emerging radiation when a non-zero torque value is employed at the inner edge of the disk.

The outline of the paper is as follows. In Sec.~\ref{sec2}, we introduce the naked singularity and black hole spacetimes. Then we obtain the physical parameters of the circular orbits such as specific energy, specific angular momentum, and angular velocity in static and spherically symmetric spacetimes in section~\ref{sec3}. In Sec.~\ref{sec4}, we review the physical properties of the Novikov-Thorne accretion disks and obtain the equation of flux for the spacetimes with the matching hypersurface. The radiating flux, spectral distribution, accretion efficiency, and temperature distribution for black hole and naked singularity spacetimes are discussed in Sec.~\ref{sec5}. We conclude our results in Sec.~\ref{sec6}. We choose the system of units such that $c=h=G=\sigma=k=1$, where $c$ is the speed of light, $h$ is the Planck's constant, $G$ is the Newton's gravitational constant, $\sigma$ is the Stefen-Boltzmann constant, and $k$ is the Boltzmann constant. Here we use $(-,+,+,+)$ sign convention.
\section{Spacetime geometries of black holes and naked singularities}\label{sec2}
\subsection{Null naked singularity spacetime (NNS)}
The metric of the static and spherically symmetric NNS spacetime can be written as~\cite{Joshi:2020tlq},
\begin{equation}
    ds^2 = -\left(1+{\frac{M_{T}}{r}}\right)^{-2}dt^2 + \left(1+{\frac{M_{T}}{r}}\right)^{2} dr^2
    + r^2d\Omega^2\, ,
    \label{nullmetric}
\end{equation}
where $d\Omega^2= d\theta^2 + \sin^2\theta d\phi^2$ and $M_{T}$ is the ADM mass of the spacetime. It is shown in \cite{Joshi:2020tlq} that there is a strong curvature singularity at the centre of the spacetime in Eq.~(\ref{nullmetric}). There is no event horizon around the singularity, and hence the central singularity is naked. The singularity at $r=0$ is nulllike, and therefore, we call this spacetime nulllike naked singularity spacetime.  At a large distance from the center, the NNS spacetime in Eq.~(\ref{nullmetric}) starts to mimic the Schwarzschild metric. Moreover, one can verify that the spacetime satisfies the weak, null, and strong energy conditions~\cite{Joshi:2020tlq}.
\subsection{JMN1 naked singularity spacetime}
The JMN1 spacetime can be formed as an equilibrium configuration of gravitational collapse of an anisotropic matter fluid.
Its metric is given by the following way~\cite{Joshi:2011zm}
\begin{equation}
    ds^2 = -(1-M_0)\left(\frac{r}{r_b}\right)^{M_0/(1-M_0)}dt^2 + \frac{dr^2}{1-M_0} + r^2d\Omega^2\, .
    \label{jmn1metric}
\end{equation}
Here, $M_0$ is a positive constant which is always less than $1$, and $r_{b}$ is the boundary radius. At $r=0$ and $M_{0}<\frac{2}{3}$, the spacetime has a timelike naked singularity. Since the JMN1 spacetime is not asymptotically flat, it is matched to the Schwarzschild spacetime at the matching radius $r=r_b$. The spacetime geometry of the Schwarzschild black hole is described as
\begin{equation}
    ds^2 = -\left(1-\frac{2M_T}{r}\right)dt^2 + \frac{dr^2}{1-\frac{2M_T}{r}} + r^2 d\Omega^2 .\label{sch_metric}
\end{equation}
The total mass $M_T$, $r_b$, and $M_0$ are related by the equation
\begin{equation}
    M_T=\frac{1}{2}M_0\, r_b\,.
\end{equation}
\begin{figure*}[t]
    \centering
   \subfigure[]
        {
  		\includegraphics[width=0.45\linewidth]{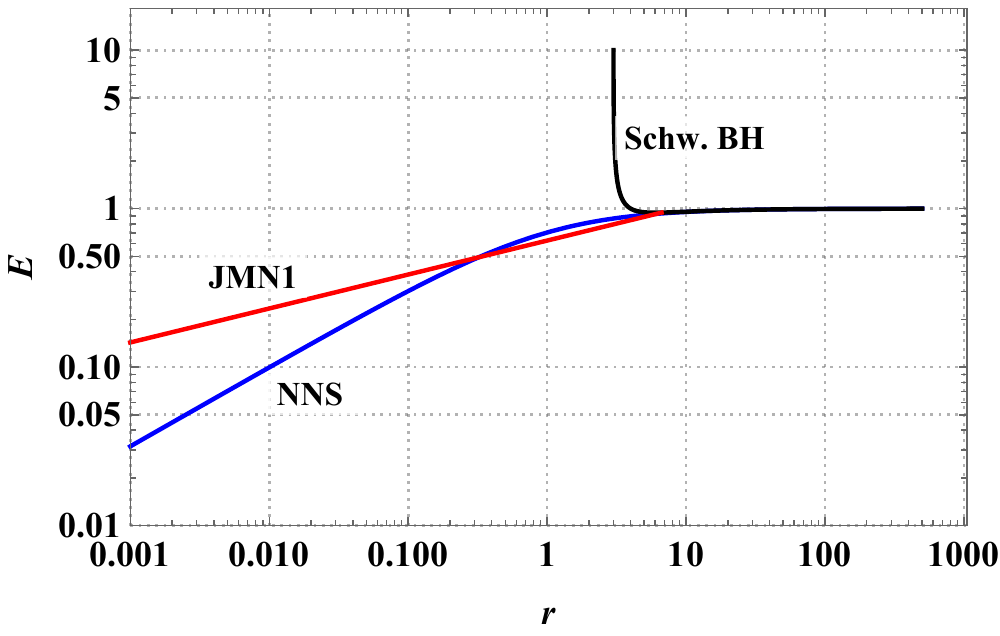} \label{fig:En}
  		}
  	\hspace{0.3cm}
    	\subfigure[]
  	    {
  		\includegraphics[width=0.45\linewidth]{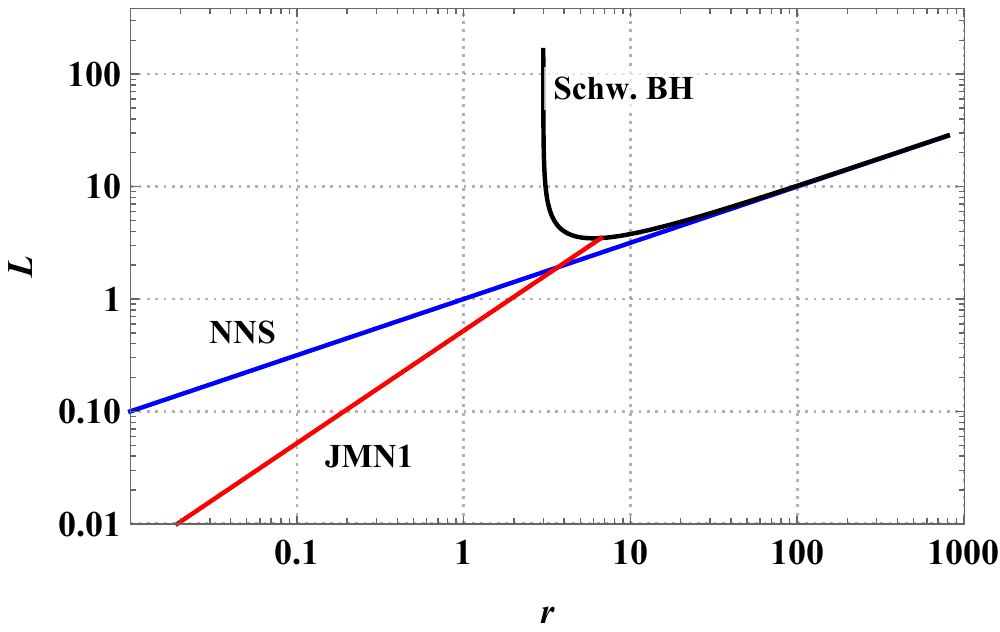}
        \label{fig:L}
  	    }
  	\hspace{0.3cm} 
  	\\
  	
  	\subfigure[]
  	    {
  		\includegraphics[width=0.45\linewidth]{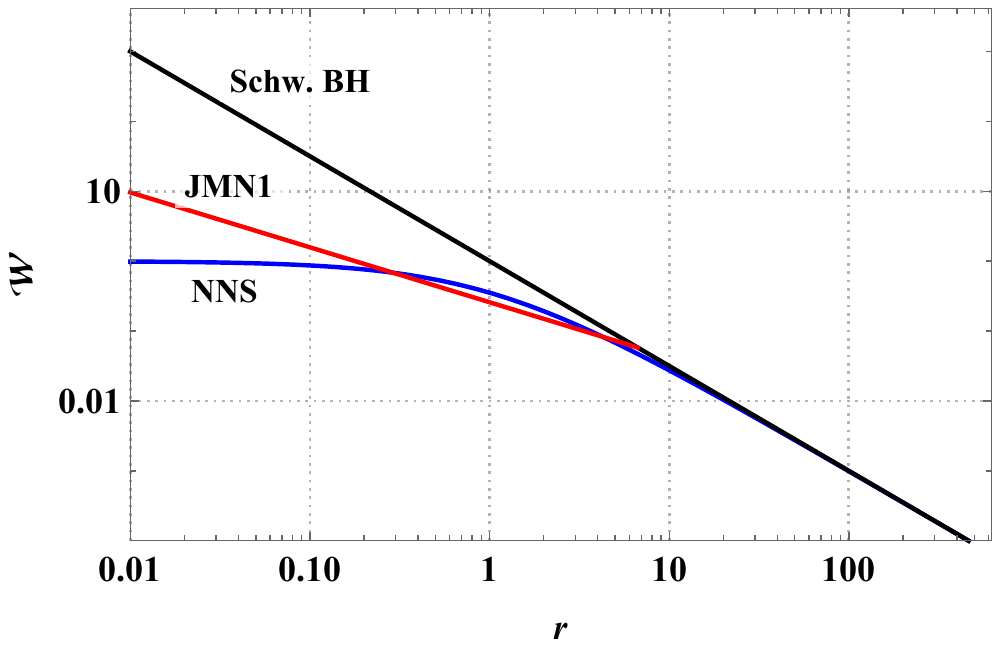}
  		\label{fig:omega}
  		}
  		
  	  \caption{The orbital properties specific energy $E$, specific angular momentum $L$, and angular velocity $\mathcal{W}$ for the Schwarzschild (black), NNS (blue) and JMN1 (red) spacetimes for $M_T=1$. For JMN1, we consider $M_{0} = 0.3$ and $r_{b} = 6.667$.} 
  \label{fig1}
\end{figure*}
\section{Circular timelike geodesics in spherically symmetric spacetimes}\label{sec3}
The line element for a general spherically symmetric and static spacetime is given by
\begin{equation}
    ds^2 = - A(r)dt^2 + B(r)dr^2 + C(r)d\Omega^2\,\,.
    \label{metric}
\end{equation}
Due to the spherical symmetry of spacetime, we can consider $\theta=\pi/2$, ensuring that the test particle moves along the timelike geodesic in the equatorial plane. Since spacetime has temporal and spherical symmetries, the energy per unit rest mass ($E$) and the angular momentum per unit rest mass ($L$), respectively, of the particle orbiting the compact object are conserved along the geodesic. These are defined as
\begin{equation}
    E = A(r)\,\dot{t}\,,\,\,\,\,\, L = C(r)\,\dot{\phi}\,.
\end{equation}
Here, an overdot represents the derivative with respect to the proper time, $\tau$. Using above equations and the timelike geodesics condition ($g_{\mu\nu}u^\mu u^\nu = -1$) for equatorial plane, we can write,
\begin{equation}
    AB\,\dot{r}^2 + V_{eff}(r) = E^2,
\end{equation}
and the effective potential $V_{eff}(r)$ is defined as
\begin{equation}
    V_{eff}(r) = \left(1+\frac{L^2}{C(r)}\right)A(r).
\end{equation}
For stable circular orbits in the equatorial plane, the condition $V_{eff} = E^2$, $V'_{eff} = 0$ and $V''_{eff}>0$ must hold. Here the $(')$ denotes the differentiation with respect to the coordinate $r$. From the first two conditions, it is easy to obtain
\begin{eqnarray}
    E &=& \frac{A}{\sqrt{A - C \mathcal{W}^2}},\\
    L &=& \frac{C\mathcal{W}}{\sqrt{A - C \mathcal{W}^2}},\\
    \mathcal{W} &=& \sqrt{\frac{A'}{C'}}\,,
\end{eqnarray}
where, $\mathcal{W} = d\phi/dt$ is the orbital angular velocity of the test particle. The radius of the marginally stable circular orbit $r_{ms}$ (or innermost stable circular orbit) of the particle orbiting the compact object can be found out using the condition
\begin{equation}
    V''_{eff} = 0.\label{v_dbl_prime}
\end{equation}
The marginally stable orbit is the innermost stable circular orbit in which a particle can stably revolve around a massive object.

Using the conditions for circular orbits and Eq.~(\ref{v_dbl_prime}),  we find that the marginally stable circular orbit is located at $r_{ms}=6 M_T$ for the Schwarzschild black hole. Similarly, we confirm that NNS and JMN1 spacetimes do not have an innermost stable circular orbit; therefore, circular orbits are possible from large radii up to $r \to 0$.

The comparative study of the properties of stable circular orbits $E$, $L$, and $\mathcal{W}$ for the NSS, JMN1, and Schwarzschild spacetimes is shown in Fig.~(\ref {fig1}). The matching boundary of JMN1 is set to $r_b = 6.667$, where it matches with the exterior Schwarzschild spacetime. The total mass $M_T$ is set to $1$ throughout the paper. Figs.~(\ref{fig:En}) and~(\ref{fig:L}) represent the energy per unit rest mass $E$ and angular momentum per unit rest mass $L$ as a function of radius $r$. In NNS and JMN1, both $E$ and $L$ go to zero as $r\to0$. The behaviour of $E$ and $L$ in NNS and JMN1 at a large distance from the centre resembles that of the Schwarzschild black hole. For the Schwarzschild black hole, the lowest values of $E$ and $L$ are at $r_{ms}= 6 M_T$, and below $r=6 M_T$, $E$ and $L$ increase as $r$ decreases, which is the result of the unavailability of stable circular orbits. 

The corresponding results for angular velocity $\mathcal{W}$ are shown in Fig.~(\ref{fig:omega}). The distinguished behaviour of circular orbit properties in naked singularities is also expected to be reflected in the spectral properties of their accretion disks, which are discussed later in this paper.
\section{Physical properties of thin accretion disks}\label{sec4}
We consider the geometrically thin and optically thick accretion disk model described by Novikov-Thorne~\cite{Novikov, Page:1974he}. In this section, we first briefly review some of the physical properties of the Novikov-Thorne accretion disk obtained in~\cite{Page:1974he}. Then we obtain the equation for energy flux for the case when the interior spacetime is matched with the exterior spacetime at a matching radius.

The massive particles move in nearly circular orbits in the accretion disk. For naked singularities considered here, we assume that the test particle orbiting inside the matter cloud is influenced only by the cloud's gravity and has no interaction with the matter cloud that constitutes naked singularity spacetime. The accretion disk is optically thick if the free mean path of the scattering photons $l$ is much smaller than the height of the disk $H$, described by the maximum half-thickness of the disk, $l<<H$~\cite{2017bhlt.book.....B}. Here, we consider a thin accretion disk whose vertical size is negligible in comparison to its horizontal extension, i.e., the disk height $H$, is considerably less than its characteristic radius $R$, $H<<R$~\cite{Novikov, Page:1974he}. This means that most of the matter lies in the equatorial plane. In this case, the vertical size specified along the z-axis in cylindrical coordinates $(r,\phi, z)$ is insignificant. The efficient cooling provided by radiation across the disk surface keeps the disk from accumulating heat produced by stresses and dynamical friction. Due to this cooling process, the disk remains in hydrodynamic equilibrium. In a steady-state accretion disk model, the mass accretion rate, $\dot{m}$, is considered to be constant and does not vary with time. The inner edge of the accretion disk is at a marginally stable circular orbit $r_{ms}$, and the accreting matter at the higher radii follows Keplerian motion. The physical characteristics defining the accreting matter are averaged over a specific time scale, such as $\Delta t$, the entire period of the orbits, the azimuthal angle $\Delta \phi =2\pi$, and the disk height $H$. The particles in the accretion disk move with a rotational velocity $\mathcal{W}$, specific energy $E$, and specific angular momentum $L$ that depends entirely on the radial coordinate $r$. The accreting particles orbit with the four-velocity $u^{\mu}$ and constitute the disk of average surface density $\Sigma$, which is described as
\begin{equation}
    \Sigma = \int_{-H}^{H} <\rho_0> dz\,.
\end{equation}
Here $\rho_0$ represents the rest mass density, and $<\rho_0>$ denotes the average of rest mass density over time $\Delta t$ and angle $2 \pi$. The energy-momentum tensor describing the source of orbiting particles is an anisotropic fluid source described as
\begin{equation}
 T^{\mu \nu } = \rho_0 u^\mu u^\nu + 2 u^{(\mu} q^{\nu)} + t^{\mu \nu },
\end{equation}
where $  u_\mu q^\mu  =  0$ and $u_\mu t^{\mu \nu }  =  0$. The quantities $q^\mu$ and $ t^{\mu \nu } $ are the energy flow vector and the stress tensor, respectively, which are measured in the averaged rest-frame of the particle orbiting in the disk. Specific heat is not taken into account in this case.  One of the important properties of the time-averaged disk structure is the energy flux $ \mathcal{F}(r) $ across the disc surface, which can be calculated by averaging the energy flow vector over time and the azimuthal angle given by,
\begin{equation}
\mathcal{F}( r )  =   <q^z> .
\end{equation}
Another significant quantity of the radial structure of the disk is the time-averaged torque due to the stresses in the disk, and it is defined as
\begin{equation}
W_\phi^r = \int_{-H}^H <t_\phi^r> dz\,.
\end{equation}
Here $<t_\phi^r>$ denotes the time and orbital averaged $\phi$-$r$ component of the stress tensor. The energy and angular momentum flux 4-vectors in terms of the stress-energy tensor  $ T^{\mu \nu } $ of the disk are respectively given by
\begin{equation}
 -E^\mu = T^\mu_ \nu  \left( \frac{ \partial }{ \partial t} \right)^\nu    ~\text{and} ~ J^\mu = T^\mu_ \nu  \left( \frac{ \partial }{ \partial  \phi} \right)^\nu\,.
\end{equation}
Using the rest mass conservation law,
\begin{equation}
\nabla_\mu (\rho_0u^\mu) =0,
\end{equation}
one can obtain the time-averaged rest mass accretion rate that does not depend on the disk radius,
\begin{equation}
\dot{m} \equiv - 2\pi \sqrt{-g}~ \Sigma ~u^r = ~\text{constant},
\end{equation}
where  an overdot denotes  the derivative with respect to the time  coordinate  ($t$), $g$ is the determinant of a metric tensor. From the energy conservation,  $\nabla_\mu E^\mu =0 $, one can find,
\begin{equation}
[\dot{m} E - 2 \pi \sqrt{-g}\,\, \mathcal{W}~\,W_\phi^r ]_{,r} = 4 \pi \sqrt{-g} \mathcal{F} E,\label{E_conservation}
\end{equation}
where $(_{,r})$ means  the differentiation with respect to the radial coordinate $r$. The Eq.~(\ref{E_conservation}) implies that the energy carried by the rest mass of the disk, $\dot{m} E$, and the energy carried by the torques in the disk, $2 \pi \sqrt{-g}\, \mathcal{W} \,W_\phi^r$, are  well-balanced by the energy radiated away from the disk's surface, $4 \pi \sqrt{-g}\,\mathcal{F} E$.
Using the law of angular momentum conservation,  $\nabla_\mu J^\mu =0 $, one can write,
\begin{equation}
[\dot{m} L - 2 \pi \sqrt{-g}  ~W_\phi^r ]_{,r} = 4 \pi \sqrt{-g} \mathcal{F} L\,. \label{L_consevation}
\end{equation}
The first term on the left-hand side of Eq.~(\ref{L_consevation}) represents the angular momentum transported by the rest mass of the disk; the second term denotes the angular momentum carried by the torques in the disk; and the third term on the right-hand side represents the angular momentum transferred from the surface of the accretion disk via outgoing radiation. For the ease of calculation, change variables to~\cite{Page:1974he}
\begin{eqnarray}
    f & = & 4 \pi\sqrt{-g} ~\mathcal{F}/\dot{m},\\
    w &= &2\pi \sqrt{-g} ~\,W_\phi^r / \dot{m}\,.\label{w1}
\end{eqnarray}
In terms of $f$ and $w$, the conservation laws (\ref{E_conservation}) and (\ref{L_consevation}) become
 \begin{eqnarray}
     (L-w)_{,r}&=&f L\,,\label{fL}\\
   (E-\mathcal{W} \,w)_{,r}&=&f E\,.  \label{fE}
 \end{eqnarray}
 Using above equations along with universal energy-angular-momentum relation $dE=\mathcal{W} \,dL$, we can write
 \begin{equation}
     \textit{w}=[(E-\mathcal{W} L)/(-\mathcal{W}_{,r})] f\,.\label{w2}
 \end{equation}
 Eliminating $w$ from Eq.~(\ref{fL}) using above expression, and integrating the resulting first-order differential equation for $f$, we get
 \begin{equation}
    \frac{(E-\mathcal{W} L)^2}{-\mathcal{W}_{,r}}f= \int(E-\mathcal{W} L)L_{,r}dr + \text{const}.\label{f}
 \end{equation}
   Here, the constant of integration was determined considering the physical fact that accreting matter falls off the disk when reaches the innermost stable circular orbit, $r=r_{ms}$, and it spontaneously drops into the centre of the compact object. As a result, there is insignificant matter just within $r=r_{ms}$ to induce torque on the matter just outside $r=r_{ms}$.  This implies that the torque  $W^r_{\phi}$, and hence $f(r)$, must disappear at $r=r_{ms}$. To make $f(r_{ms})$ vanish, the integration constant must be selected in such a way that
    \begin{equation}
        \frac{(E-\mathcal{W}  L)^2 }{ -\mathcal{W}_{,r}} f= \int_{r_{ms}}^{r} (E-\mathcal{W}  L)\,L_{,r} \, dr\,.
    \end{equation}
  Therefore the expression of time-averaged flux radiating from the disk surface can be written as~\cite{Novikov, Page:1974he}, 
  \begin{equation}
        \mathcal{F}(r)=-\frac{\dot{m}}{4\pi \sqrt{-g}}\frac{\mathcal{W}_{,r}} {(E-\mathcal{W} L)^2}\int^r_{r_{\rm ms}}(E-\mathcal{W} L)L_{,r}dr.
        \label{flux} 
  \end{equation}
It should be noted that this expression of flux (Eq.~(\ref{flux})) can only be employed for asymptotically flat spacetimes. For non-continuous spacetimes, where there is a matching between the interior and exterior spacetimes at a particular hypersurface, the boundary conditions in Eq.~(\ref{f}) change at the matching hypersurface. In such cases,  Eq.~(\ref{flux}) needs to be modified depending on whether the emitting point is in the interior or exterior spacetime.
\subsection{Equation for the energy flux for spacetimes with matching boundary}
\begin{figure}[h]
    \centering
  	\includegraphics[width=0.9\linewidth]{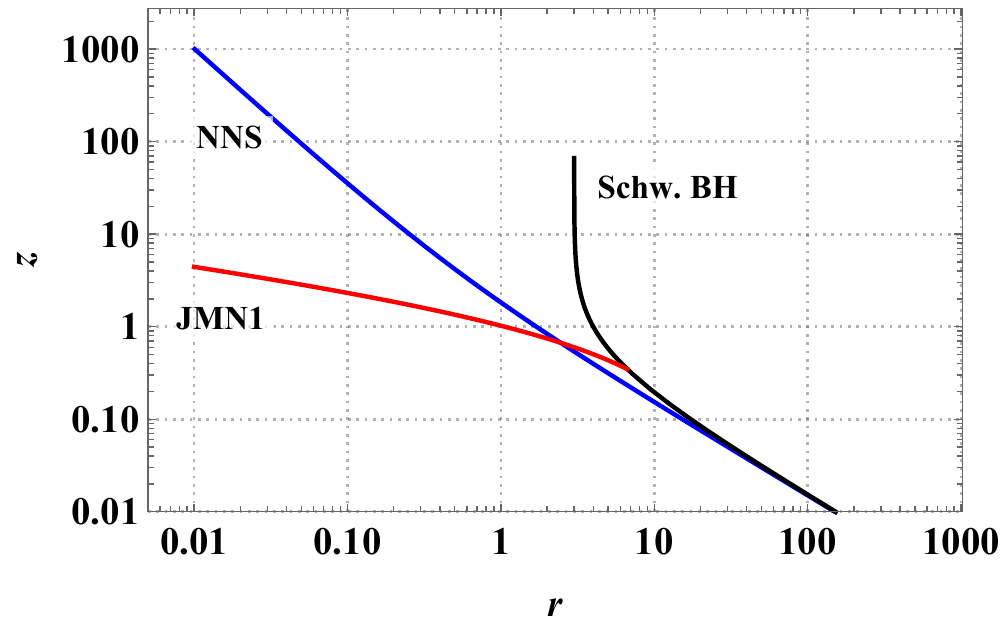}
     \caption{The characteristic redshift $(z)$ vs $r$ for a black hole (black), NNS (blue) and JMN1 (red) spacetimes. The figure shows how the photons emitted from the disk surface are modified due to both gravitational redshift and Doppler shift before reaching the distant observer. For Schwarzschild black hole, $z$ tends to infinity at $r = 3 M_T$. On the other hand, for NNS and JMN1, $z$ dominates at $r \to 0$ and hence $z\to \infty$ at $r \to 0$. The total mass $M_T$ is set to $1$.}
  \label{fig:redshift}
\end{figure}

In this subsection, we obtain the equation for the radiating flux for spacetimes with matching boundaries.  We consider that a timelike hypersurface divides the spacetime into two regions with interior and exterior spacetime metrics $g^{-}_{\alpha \beta}$ and $g^{+}_{\alpha \beta}$, respectively. In general relativity, for the smooth joining of the interior and exterior metrics at the junction, the induced metrics~($h_{ab}$) on both the sides of the matching hypersurface must be identical~\cite{E_poisson}. Therefore, 
   \begin{equation}
   h_{ab|int} = h_{ab|ext}
   \end{equation}
  where  $h_{ab|int}$  and $h_{ab|ext}$ are, respectively, the induced metrics of internal and external spacetimes on the matching boundary. We assume that the matching boundary of the internal and external spacetimes is at $r=r_b$. Here we derive the expression of flux for the scenario where $ r_b $ is greater than the radius of the marginally stable orbit of the exterior spacetime~($r_{2ms}$). In this instance, there would be a single continuous accretion disk spanning from a specific outer radius in the exterior spacetime to the marginally stable orbit of the interior spacetime~($r_{1ms}$). For simplicity, we define
   \begin{equation}
     \Phi(r)=\frac{(E-\mathcal{W}  L)^2}{-\mathcal{W}_{,r} }\,\hspace{5pt} \text{and} \hspace{5pt}\Psi(r)=  \int (E-\mathcal{W}  L) L_{,r} \, dr.\\
   \end{equation}
 Hence, for internal spacetime $g^{-}_{\alpha \beta}$ $(r<r_b)$, Eq.~(\ref{f}) in terms of $\Phi(r)$ and $\Psi(r)$ can be written as
    \begin{equation}
    \Phi_1(r) f_1(r) = \Psi_1(r) + c_1\,, \label{i1}
    \end{equation}
where subscript $(_1)$ denotes the parameters corresponding to the internal spacetime, and $c_1$ is the integration constant. Now, considering the ``no-torque'' condition at the inner boundary of the disk, i.e., $f_1(r_{1ms})=0$, we get
   \begin{equation}
       \Phi_1(r)\, f_1(r) =\Psi_1(r)-\Psi_1(r_{1ms})\,. \label{i2}
   \end{equation}
    Substituting the expressions of $\Phi_1(r)$ and $\Psi_1(r)$, above equation becomes
    \begin{equation}
          \frac{(E_1-\mathcal{W}_1  L_1)^2 }{ -\mathcal{W}_{1,r}} f_1= \int_{r_{1ms}}^{r} (E_1-\mathcal{W}_1  L_1)\,L_{1,r} \, dr\,,
    \end{equation}
which is same as Eq.~(\ref{f}). Therefore, when the point of emission lies in the internal spacetime,  the flux of electromagnetic radiation can be calculated using Eq. (\ref{flux}), which we denote as $\mathcal{F}_1$.
Similarly for the external spacetime $ g^{+}_{\alpha \beta} $~( $r > r_b$ ), Eq.~(\ref{f}) becomes
    \begin{equation}
        \Phi_2(r)\, f_2(r) = \Psi_2(r) +c_2 \label{e1}\,\,
    \end{equation}
with subscript $(_2)$ denoting the parameters associated with the exterior spacetime.
Assuming that the distribution of energy flux remains continuous from interior to exterior geometry, we can employ the boundary condition that at $r=r_{b}$,\, $\mathcal{F}_1(r_b)=\mathcal{F}_2(r_b)$ to determine $c_2$. Therefore,
    \begin{equation}
         f_2(r_b) = f_1(r_b) = \frac{\Psi_1 (r_b) - \Psi_1(r_{1ms})}{\Phi_1 (r_b)}.
    \end{equation}
    Using the above boundary condition at $r=r_b$, it is straightforward to obtain from Eq.~(\ref{e1}),
    \begin{equation}
        f_2(r)=\frac{\Psi_2(r)-\Psi_2(r_b)}{\Phi_2(r)}+\frac{ \Phi_2(r_b) \left( \Psi_1(r_b) - \Psi_1(r_{1ms})\right)}{ \Phi_2(r)\,\Phi_1(r_b)}\,.\label{f2}
    \end{equation}
    Again, substituting the expressions of $\Phi(r)$ and $\Psi(r)$ for both the internal and external spacetimes in Eq.~(\ref{f2}), we arrive at the equation of radiating flux when the emitting point of radiation is in external spacetime,
   \begin{widetext}
    \begin{equation}
         \mathcal{F}_2(r)=-\frac{\dot{m}}{4\pi \sqrt{-g_2}}\left( \frac{\Phi_2(r_b)}{\Phi_1(r_b)}\frac{\mathcal{W}_{2
         ,r}} {(E_2-\mathcal{W}_2 L_2)^2}\int^{r_b}_{r_{\rm 1ms}}(E_1-\mathcal{W}_1 L_1)L_{1,r}dr +\frac{\mathcal{W}_{2,r}} {(E_2-\mathcal{W}_2 L_2)^2}\int^r_{r_{\rm b}}(E_2-\mathcal{W}_2 L_2)L_{2,r}d{r}\right) \,.
        \label{flux2} 
    \end{equation}
     \end{widetext}
  Assume that every local annulus of the disk emits like a blackbody, giving rise to the radiation emitting the Planck spectrum. The characteristic temperature $(T_{*})$ of the disk can be defined by
\begin{equation}
    \sigma T_*^4=\frac{\dot{m} c^2}{ [4\pi(GM_T/c^2)^2]},
\end{equation} 
where $\sigma$ is the Stefan-Boltzmann constant.
 Let $T_{BB}(r)$ be the blackbody temperature of electromagnetic radiation at radius $r$ in the local frame of accreting matter,
\begin{equation}
    T_{BB}(r)=[\mathcal{F}(r)]^{1/4} T_* \,.
\end{equation}
Here $\mathcal{F}(r)$ is the flux obtained in the frame of the observer moving along the accreting fluid, which cannot be measured directly by the observer at infinity. Before reaching a distant observer, this radiation undergoes gravitational redshifts due to energy loss as it climbs out of the gravitational potential well of the central compact object. It will also be altered by Doppler redshifts due to the rotation of the accretion disk, which is dependent on whether the gas in the accretion disk is moving towards or away from the observer. Therefore, the radiation received by the distant observer seems to have a different wavelength and temperature than that of the source. The transfer of the radiation to infinity also depends on the observer's position with respect to the accretion disk, i.e., the angle of inclination of the disk with respect to the observer. For simplicity, we assume that the observer lies on the axis of disk. As a result, the characteristic redshift $z$ can be expressed as
\begin{equation}
    1+z(r)=[-g_{tt}-\mathcal{W}^2 g_{\phi\phi}]^{-1/2}\, .
\end{equation}
The characteristic redshift accounts for the change in the photon frequency that occurs when it travels from the emitting material to the observer. The plot of $z$ as a function of distance $r$ is displayed in Fig.~(\ref{fig:redshift}).
The local blackbody temperature of the radiation emitted at $r$, as measured by the observer at infinity, is 
\begin{equation}
    T_{\infty}(r)=\frac{T_{BB}(r)}{1+z(r)} .
\end{equation}
Hence, the spectral luminosity distribution $\mathcal{L}_{\nu,\infty}$(energy per unit time) measured by the observer at infinity can be given by~\cite{Guo:2020tgv}
\begin{widetext}
\begin{equation}
    \nu {\cal L}_{\nu,\infty} = \frac{15}{\pi^4} \int_{r_{\rm in}}^\infty \left(\frac{d{\cal L}_\infty}{d\ln r}\right)\frac{(1+z)^4 (h\nu/kT_*)^4/\mathcal{F}}
{\exp[(1+z)(h\nu/kT_*)/{\mathcal F}^{1/4}]-1}\, d\ln r,
\label{integral_L}
\end{equation}
\end{widetext}
where $\frac{d{\cal L}_\infty}{d\ln r}$ is the differential luminosity given by~\cite{Novikov}
\begin{equation}
    \frac{d{\cal L}_\infty}{d\ln r} = 4\pi r \sqrt{-g} E \mathcal {F}.
\label{dLdr}
\end{equation}
Another significant physical property determining the attributes of the accretion disc is the accretion efficiency, which specifies how much of the rest mass is converted into radiation by the central compact object. Accretion efficiency can be obtained from how much energy the test particle loses coming from infinity to the disk's inner edge. Denoting the energy at infinity and inner boundary of the disk by $E_{\infty}$ and $E_{in}$ respectively, the efficiency can be expressed as
\begin{equation}
    \epsilon=\frac{E_{\infty}-E_{in}}{E_{\infty}}.
\end{equation}
Considering the energy of the test particle at the infinity as $E_{\infty}\approx1$, we obtain
\begin{equation}
    \epsilon=1- E_{in}\,.
\end{equation}
\section{Electromagnetic properties of accretion disk surrounding black holes and naked singularities}\label{sec5}
In this section, we investigate the electromagnetic and thermal properties of accretion disk such as energy flux, spectral luminosity distribution, efficiency, and temperature profile for black hole and naked singularity spacetimes.
\subsection{Energy flux emerging from the disk surface}
By computing $E$ and $L$ for each spacetime, one can easily obtain the respective radiation flux $\mathcal{F}(r)$. For the Schwarzschild black hole, the inner edge of the disk is located at $6M_T$. Therefore, the particles reaching radius $r_{in}=6 M_T$ will directly plunge into the central object, and the energy flux of the accretion disk will be cut off at $r_{in}=6M_T$. On the other hand, for NNS and JMN1 spacetimes, $r_{in}$ is set to zero. Hence,  the accretion disk spans from any arbitrary outer radius to the singularity $r\to0$.
\begin{figure}[h]
    \centering
  	\includegraphics[width=0.9\linewidth]{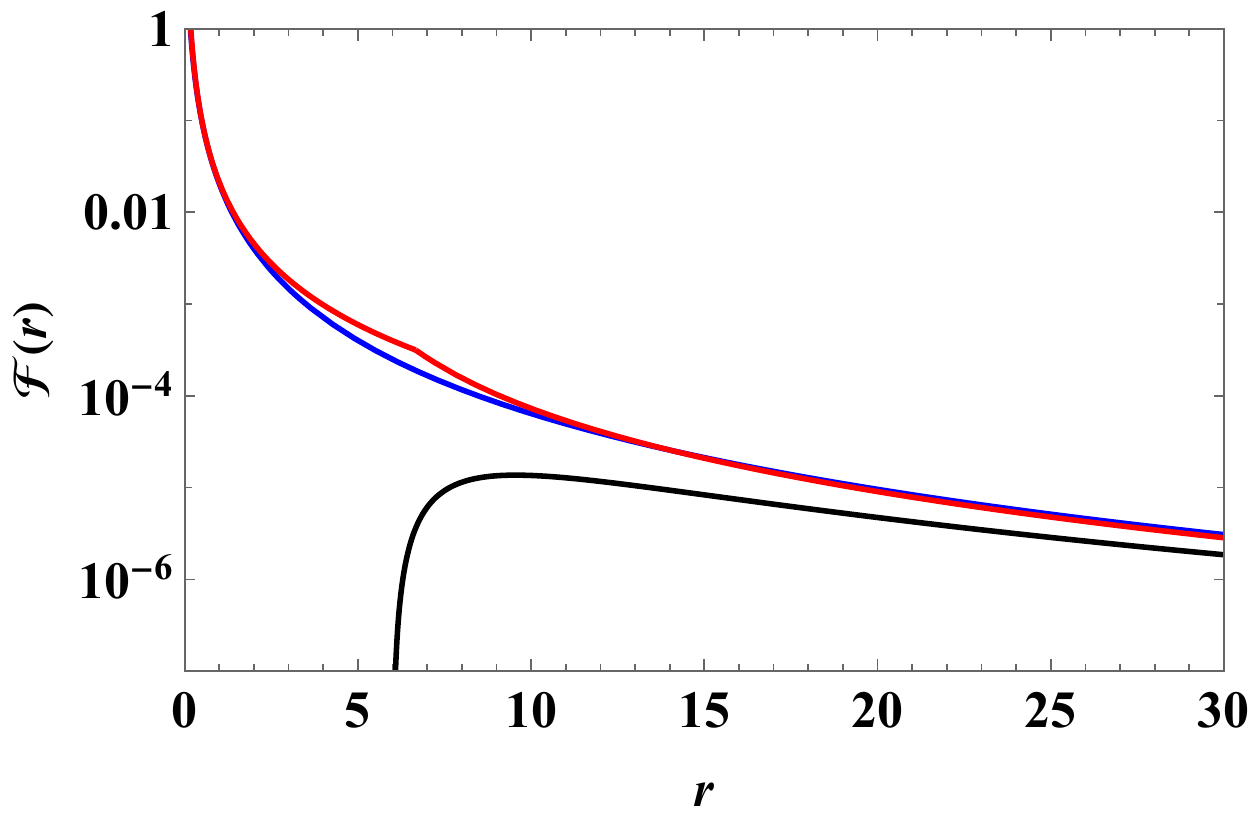}
     \caption{The figure represent the energy flux $\mathcal{F}(r)$ for a black hole (black), NNS (blue) and JMN1 (red) spacetimes as measured in local comoving frame of the accreting matter. The total mass and mass accretion rate are both set to $1$. The inner radius of the accretion disk for naked singularities is fixed at $r_{in} =10^{-7}$. For JMN1 naked singularity, we considered $M_{0}=0.3$, and $r_b=6.667$. }
  \label{fig:energy_flux}
\end{figure}
In JMN1 naked singularity model, for $M_0 > 2/3$, there is no circular orbit possible~\cite{Joshi:2011zm}. It has the matching radius $2 M_T<r_b<3M_T$ for $2/3<M_0<1$ when matched to an external Schwarzschild spacetime. Therefore, for $M_0>2/3$, there exists a single accretion disk in the external Schwarzschild geometry with an inner boundary at $r_{in}=6 M_T$. The entire JMN1 spacetime features circular orbits for $M_0<2/3$, in addition to the orbits at radii $r \geq 6 M_T$ in the external Schwarzschild spacetime.

If the matching radius $r_b$ is outside $6M_T$, then the particles in the accretion disk orbit in continuous smooth circular geodesics of external Schwarzschild solution up to $r=r_b$. Inside $r_b$, stable circular orbits are possible up to the singularity. Therefore, the particle reaching $r_b$ will then follow the series of circular orbits of the interior spacetime up to the singularity $r\to0$. Hence, there is no break in the accretion disk from the large outer radius to the singularity at the center.

Moreover, the matching radius corresponding to $1/3<M_0<2/3$ is $3 M_T< r_b<6 M_T$. Hence for $1/3<M_0< 2/3$, there are two accretion disks - (1)in the external Schwarzschild spacetime up to $r=6 M_T$, and (2)in the internal spacetime. In this case, particles approaching the marginally stable circular orbit of the external Schwarzschild spacetime plunge in with $L=L_{ms}$ till it overpasses $r_b$. Inside $r_b$, the particles settle into the circular orbit with radius $r_s$, which have the same angular momentum per unit mass as $L_{ms}$. Therefore, the outer boundary of the inner accretion disk is at $r=r_{s}<r_b$, which can be obtained by equating the specific angular momentum $L_{ms}$ with the specific angular momentum of the circular orbit of the interior spacetime. Thus, one can get
\begin{equation}
    r_{s}=r_b\sqrt{3 M_0( 2-3 M_0)}\,.
\end{equation}
The inner accretion disk spans from the singularity to $r_{s}$. On the other hand, for $0<M_0\leq1/3$, the matching radius $r_b \geq 6 M_T$. As a result, there will be a single continuous accretion disk all the way down to the singularity.

To examine the electromagnetic and thermal properties of the accretion disk around JMN1 spacetime, we analyse the case of $M_0 \leq 1/3 $. Hence, we have one continuous accretion disk with an inner part in JMN1 spacetime from $r=0$ to $r=r_b$, and an outer part in Schwarzschild spacetime up to a certain $r$.

If the point of emission of radiation lies in the internal JMN1 spacetime (i.e., $0\leq r\leq r_b$), the corresponding flux can be calculated from Eq.~(\ref{flux}). On the other hand, when the emitting point is in the external Schwarzschild spacetime (i.e., $r\geq r_b$), then we need to use Eq.~(\ref{flux2}) to obtain the energy flux $\mathcal{F}(r)$.

The flux profile for black holes and naked singularities is presented in Fig.~(\ref{fig:energy_flux}) with the black, blue, and red curves representing the flux radiated by Schwarzschild black hole, NNS, and JMN1 spacetimes, respectively. In this paper, we consider $\dot{m}=1$. It should be noted here that the singularity is not a part of spacetime, and quantum effects become dominant near the singularity. Due to this, we cannot accurately predict the thermal properties of the accretion disk up to the singularity using only general relativity. As a result, we consider the inner edge of the accretion disc at $r_{in} =10^{-7}$ when computing the physical properties of the accretion disk around naked singularities. 

From Fig.~(\ref{fig:energy_flux}), one can see that the maximum of flux of the electromagnetic radiation is higher for naked singularities as compared to the black hole. For Schwarzschild black hole, the energy flux drops to zero at $r=6$ as the accretion disk is truncated at $ r=6 $. Whereas for NNS and JMN1, $\mathcal{F}(r)\to \infty $ as the inner boundary of the disk is approached. However, in a physical scenario, the magnitude of the maximum flux must have a finite value. This indicates that the hydrodynamic and thermodynamic equilibriums are not maintained at a very small distance from the singularity. Thus, the thin disk accretion model does not provide a fair approximation near the singularity. Nevertheless, we may say that the accretion disk must be incredibly luminous close to the singularity. Aside from radiative cooling, there must be some physical processes such as energy and angular momentum transfer, as well as advection playing an essential role near the singularity.
 \begin{figure}[h!]
  	\centering
	\includegraphics[width=0.9\linewidth]{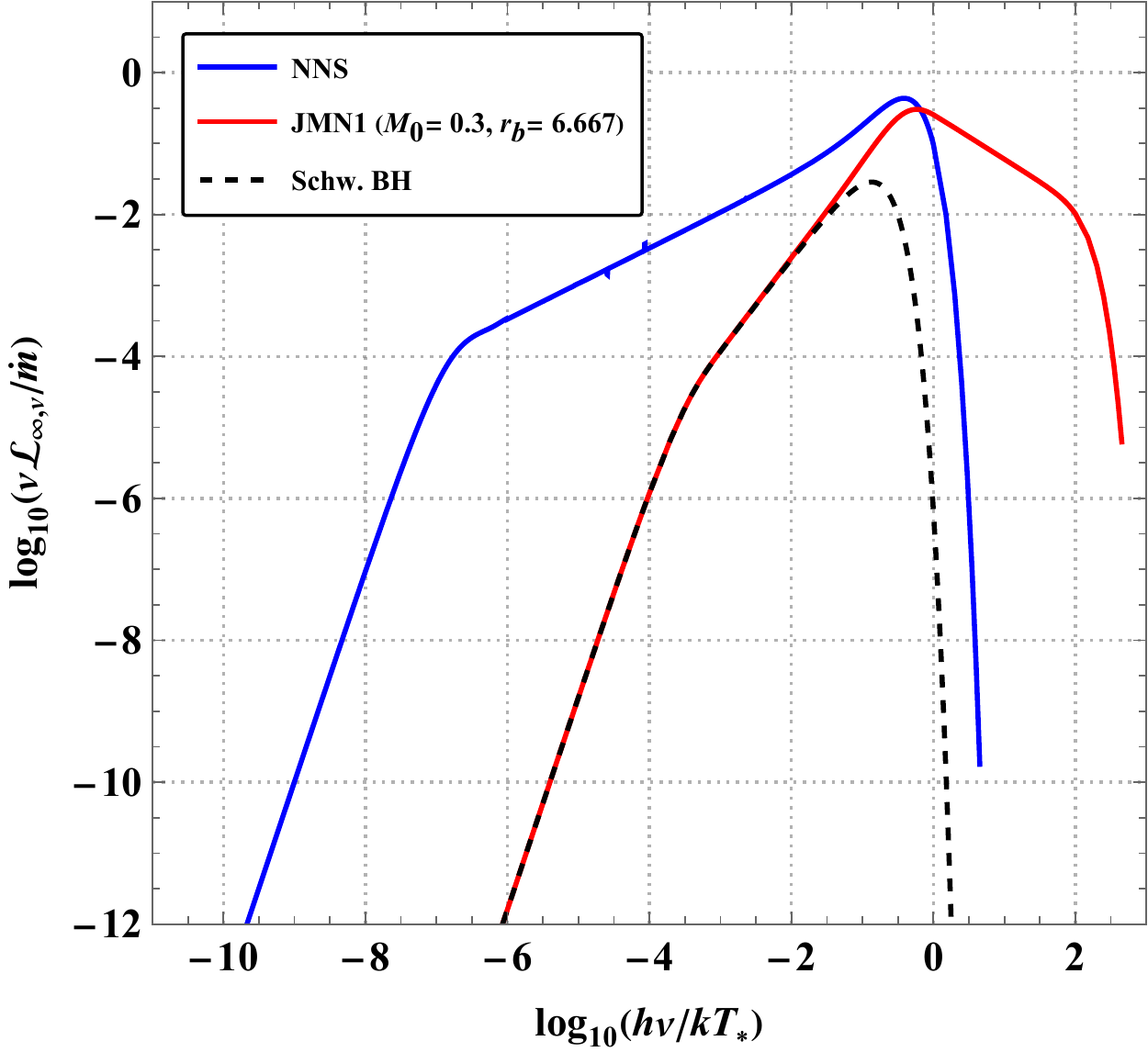}
  \caption{The Spectral luminosity distributions of radiation emitted from the accretion disks for Schwarzschild, NNS and JMN1 spacetimes.}
  \label{fig:spectra}
  \end{figure}
  \begin{figure*}[ht!]
\centering
\subfigure[NNS]
{\includegraphics[width=76mm]{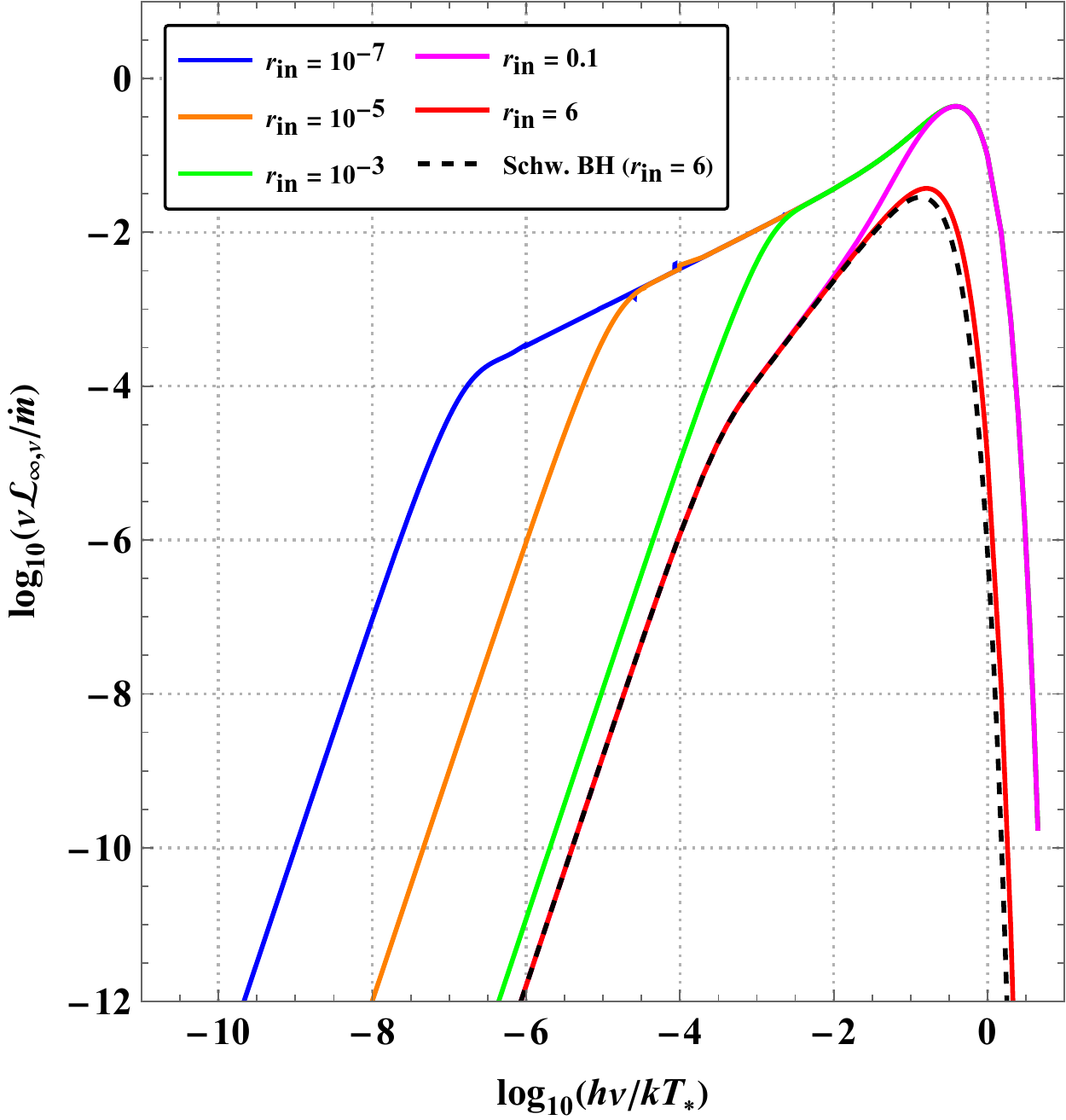}\label{fig:null_diff_rms}}
\hspace{0.5cm}
\subfigure[JMN1 naked singularity ($M_0=0.3$, $r_b=6.6667$)]
{\includegraphics[width=76mm]{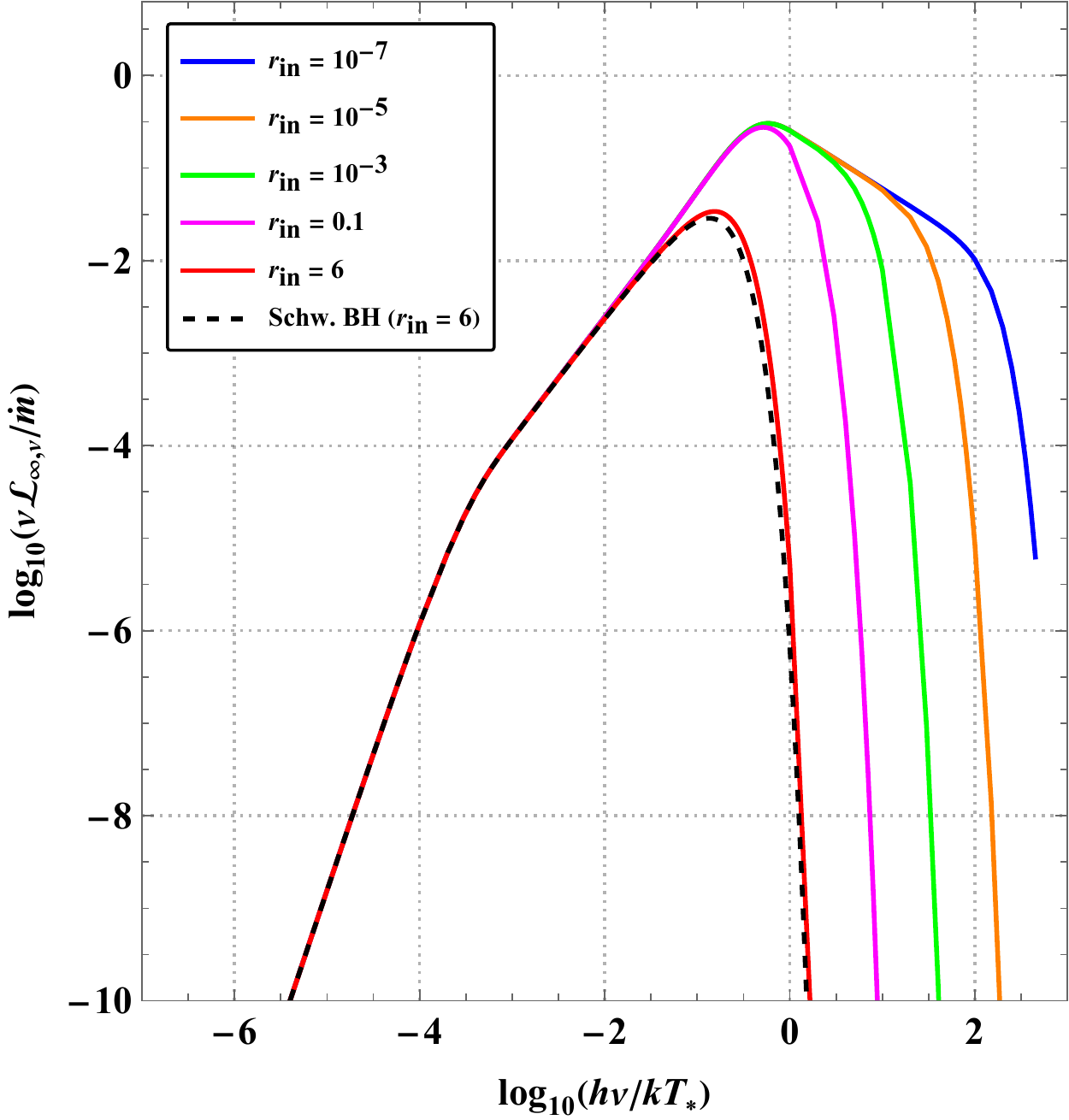}\label{fig:jmn1_diff_rms}}
 \caption{Spectral luminosity distributions of radiation emerging from the accretion disks around NNS and JMN1 spacetimes for different values of inner edge ($r_{in}$) of the disk. Here, ``zero-torque" boundary condition is assumed at $r_{in}$ to compute the energy flux. }
\label{fig:diff_rms_spectra}
\end{figure*}
\begin{figure*}[t]
\centering
\subfigure[NNS spacetime]
{\includegraphics[width=65mm]{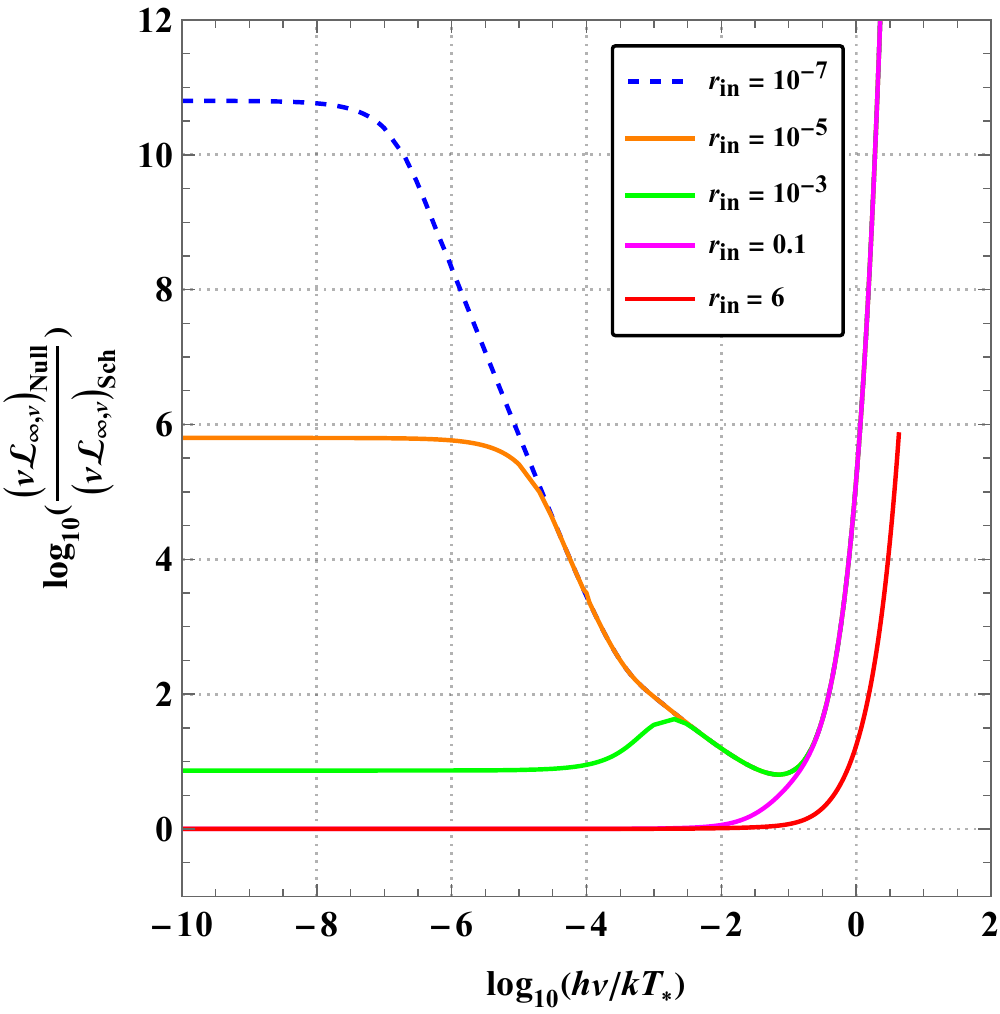}\label{fig:N_ratio}}
\hspace{0.5cm}
\subfigure[JMN1 naked singularity ($M_0=0.3$, $r_b=6.667$)]
{\includegraphics[width=65mm]{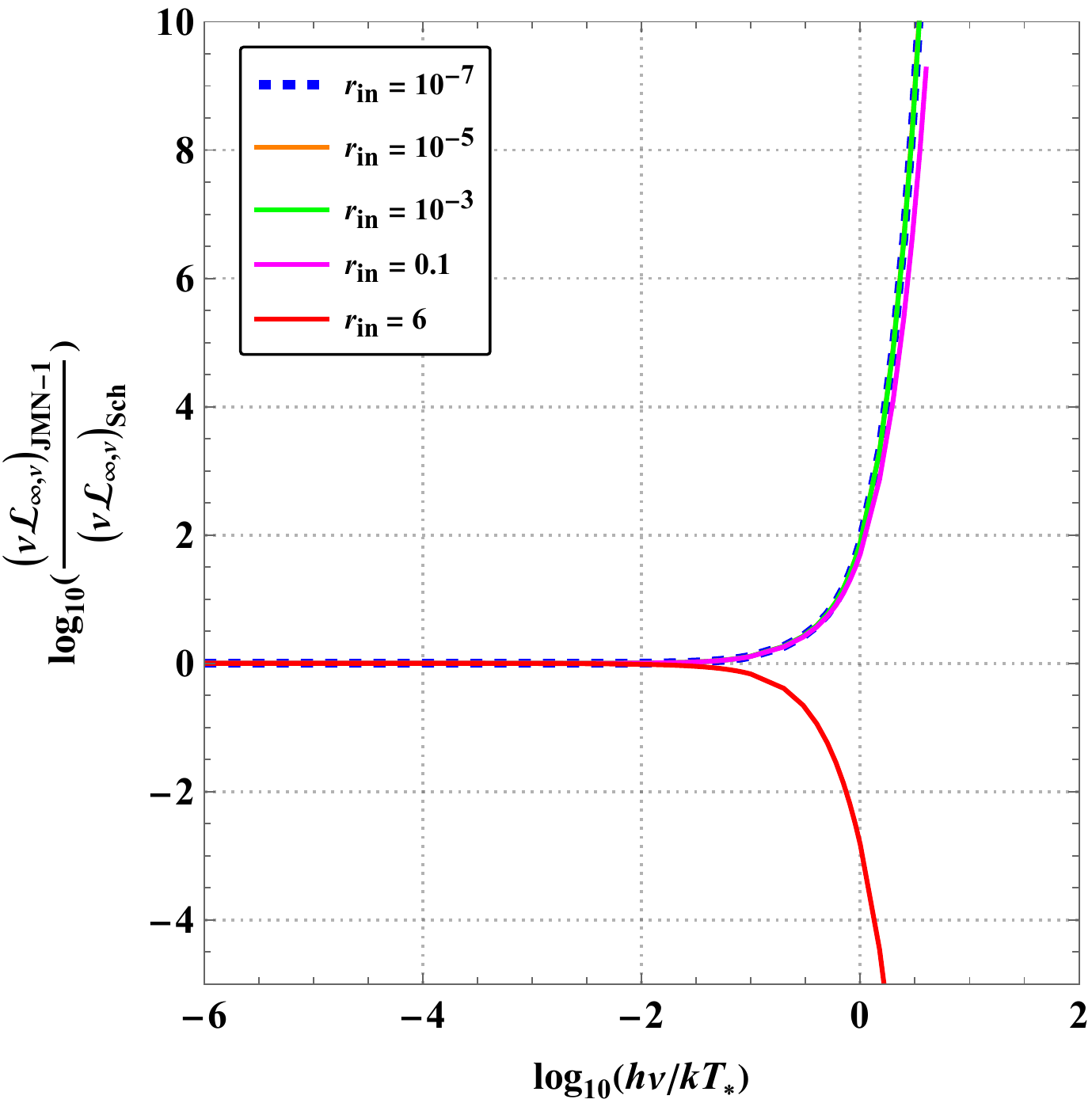}\label{fig:J_ratio}}
 \caption{The plots of luminosity ratios vs. frequency for NNS and JMN1 models for different $r_{in}$ values as considered in Fig.~(\ref{fig:diff_rms_spectra}).}
\label{fig2}
\end{figure*}
\subsection{Spectral luminosity distribution}
We integrate numerically Eq.~(\ref{integral_L}) to get the spectral luminosity distribution for the accretion disk around the black hole and naked singularity models. For the JMN1 and NNS model, the inner and outer limits of integration for Eq.~(\ref{integral_L}) are considered $10^{-7}$ and $10^{5}$, respectively. The results are represented in Fig.~(\ref{fig:spectra}). Note that here we compute the luminosity by considering the ``zero-torque" condition at the disk's inner edge.
At low frequencies, for a given mass accretion rate $\dot{m}$, JMN1 naked singularity and Schwarzschild black hole show the identical spectra. However, at high frequencies, the spectral luminosity is much higher for JMN1 naked singularity case as compared to Schwarzschild black hole. This happens as the disk in the naked singularity extends up to the singularity at the center, while for the Schwarzschild spacetime, it is truncated at the innermost stable circular orbit. For NNS, the spectrum's shape starts differing significantly from the other two models as the frequency decreases. In addition to the highly luminous spectra at higher frequencies, we receive a highly luminous spectra at lower frequencies also. 
\begin{figure*}
\centering
\subfigure[NNS spacetime]
{\includegraphics[width=76mm]{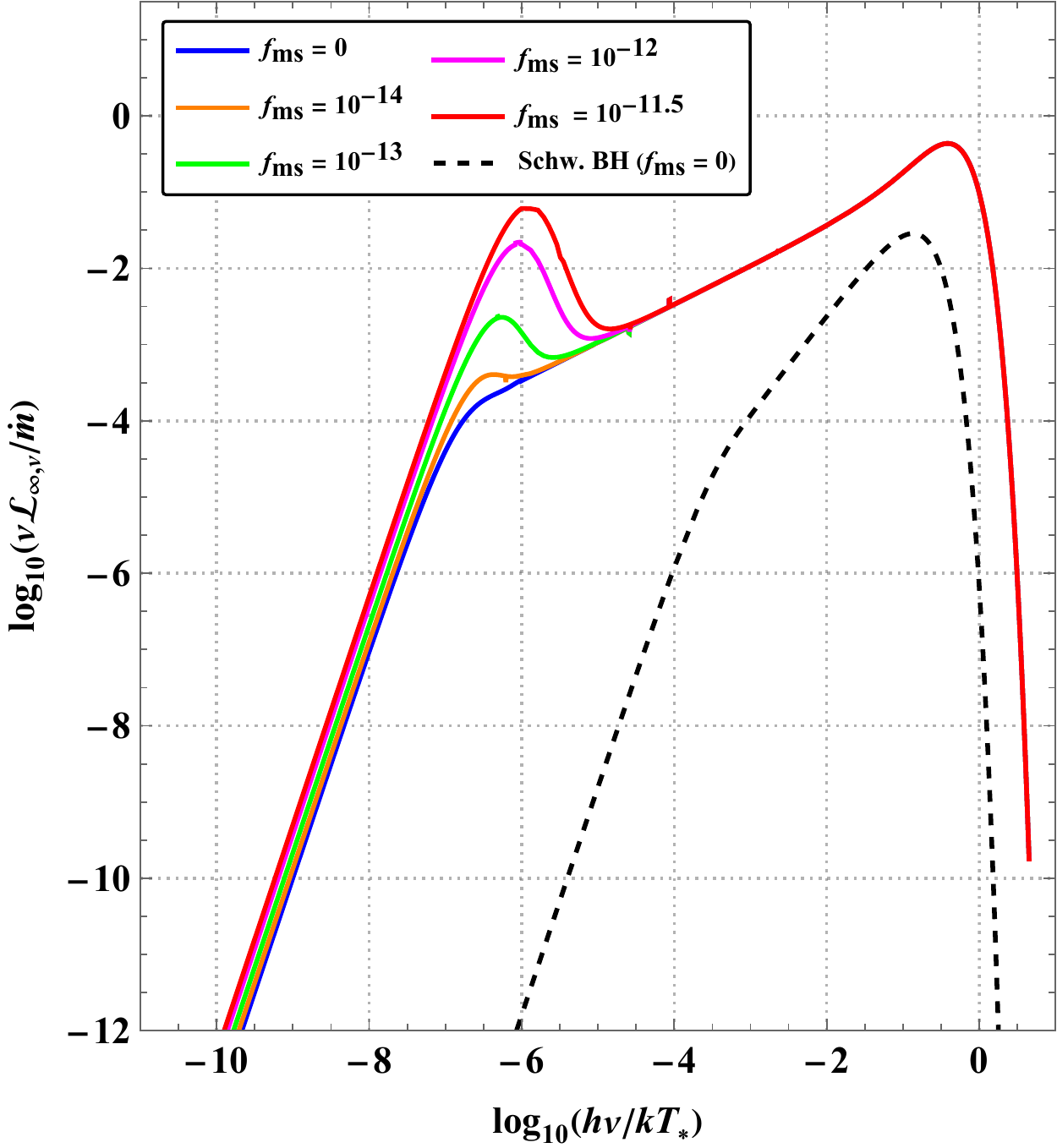}\label{fig:null_diff_tq}}
\hspace{0.5cm}
\subfigure[JMN1 naked singularity ($M_0=0.3$, $r_b=6.6667$)]
{\includegraphics[width=76mm]{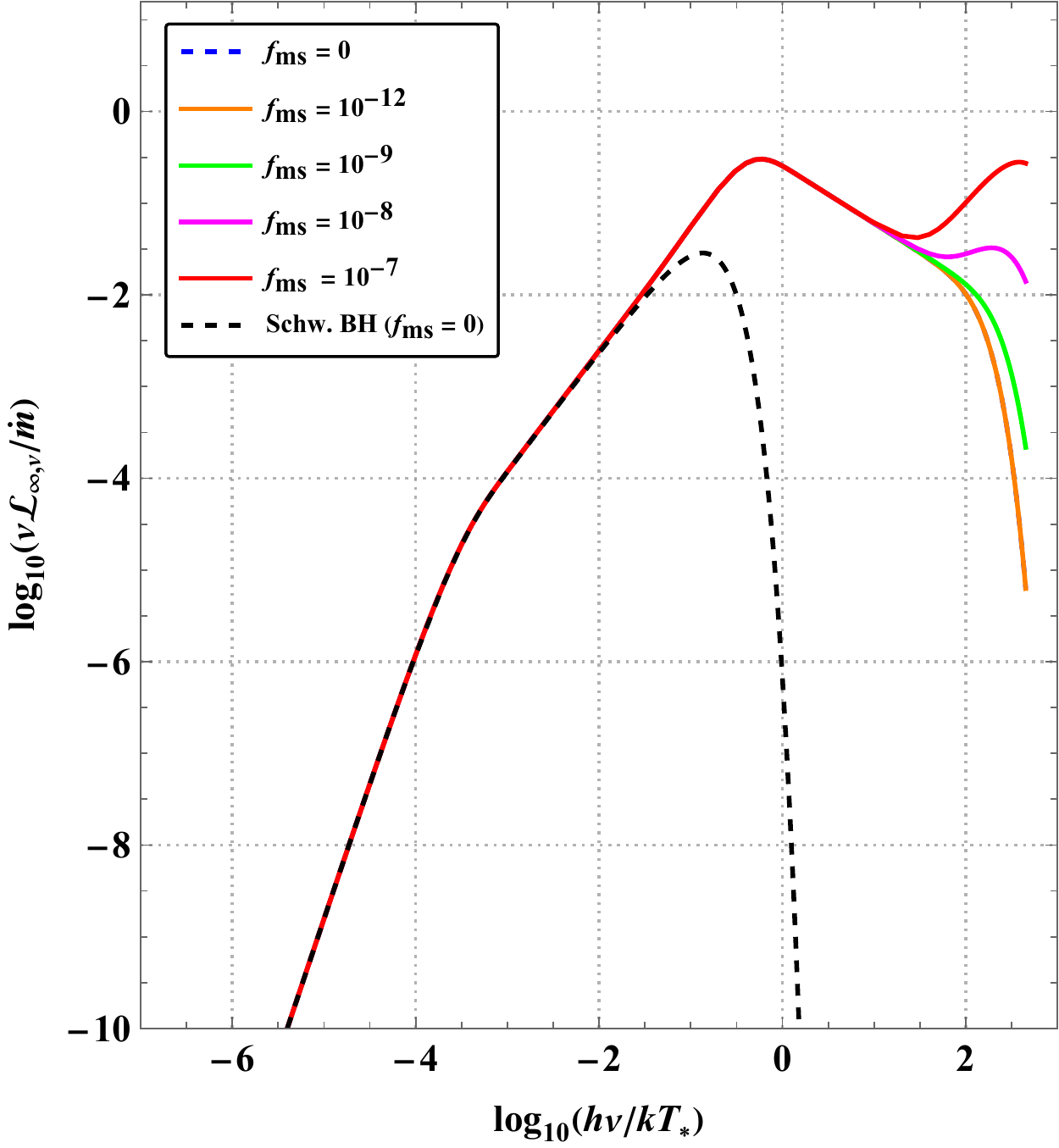}\label{fig:jmn1_diff_tq}}
 \caption{Spectral luminosity distributions of radiation from the accretion disks for NNS and JMN1 spacetimes by considering non-zero torque at the inner radius of a naked singularity. When torque is not zero at $r_{ms}$, the flux can be calculated using Eq.~(\ref{flux_fms}) by considering a non-zero value of parameter $f_{ms}$. This figure shows how spectral luminosity changes for various non-zero values of $f_{ms}$. Here, $r_{in}$ is set to $10^{-7}$ for NNS and JMN1 spacetimes.}
\label{fig:diff_tq_spectra}
\end{figure*}

To determine whether these highly luminous spectra at low frequency originate from the inner or outer part of the accretion disk, let us analyse the influence of the inner edge of the accretion disk on the luminosity spectrum in the following way. We shift the inner boundary $r_{in}$ of the disk from $10^{-7}$ to various larger radii and depict the accompanying luminosity spectra. Fig.~(\ref{fig:diff_rms_spectra}) shows these plots for NNS and JMN1 spacetimes along with the comparison with the Schwarzschild black hole. Fig.~(\ref{fig:null_diff_rms}) demonstrates that as $r_{in}$ of the accretion disk around NNS is moved away from the centre, the contribution of low-frequency radiation to producing high luminosity spectra decreases. This implies that the spectra of high luminosity at low frequencies originate in the vicinity of the singularity. This distinct characteristic of this model can help theoretically distinguish it from the disks surrounding the other two models. 

Similarly, the spectra of radiation emitted by the accretion disk of JMN1 model for different $r_{in}$ is shown in Fig.~(\ref{fig:jmn1_diff_rms}). In this figure, we observe that the luminosity spectrum shifts towards the lower frequency as the inner edge of the accretion disk is moved away from the singularity. At $r_{in}=6$, the luminosity spectra of JMN1 almost coincides with that of the Schwarzschild black hole. These results indicate that the high luminosity spectra in JMN1 is emitted by the inner regions of the accretion disk. 
  \begin{figure*}[t]
\centering
\subfigure[Schwarzschild black hole]
{\includegraphics[width=59mm]{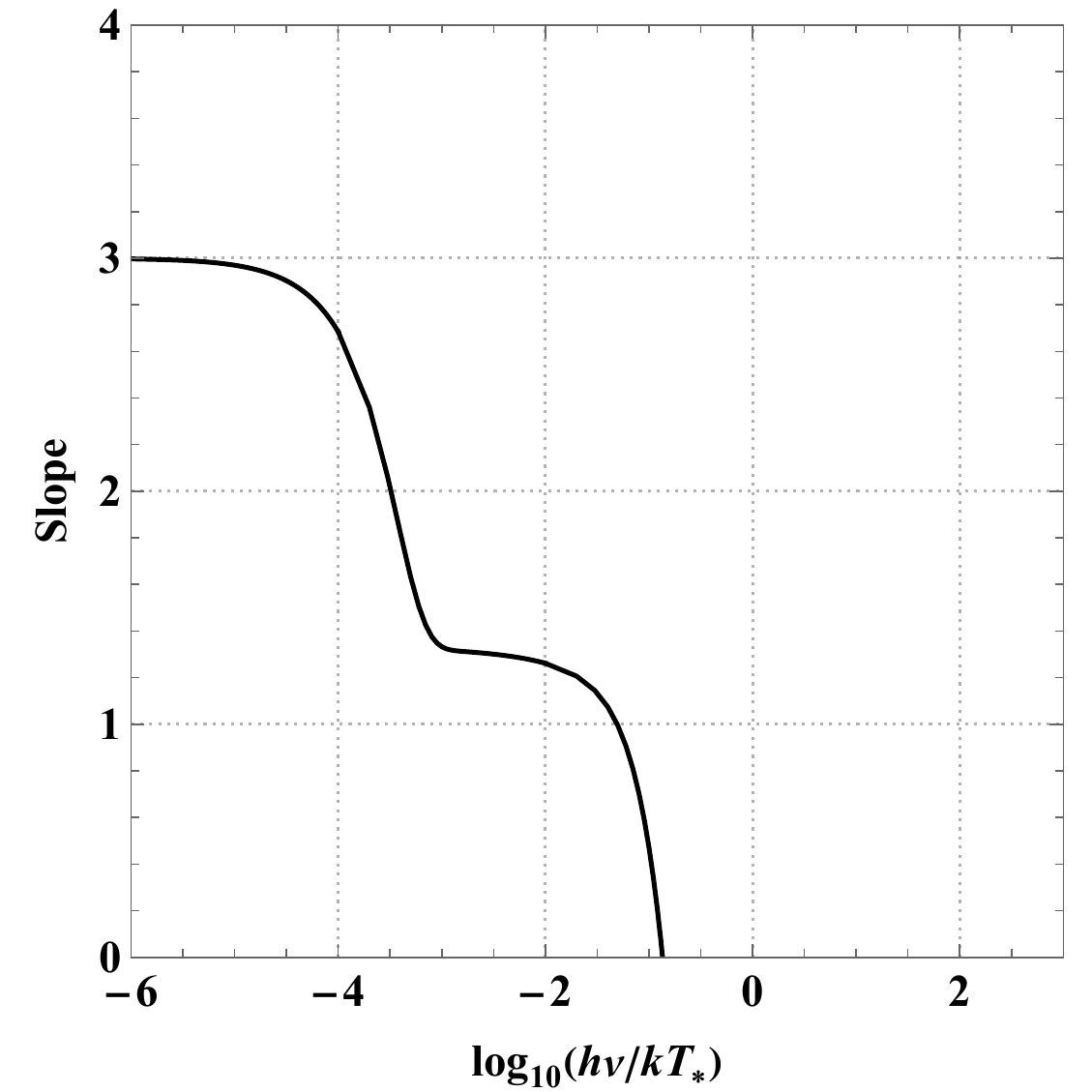}\label{fig:sch_slope}}
\subfigure[NNS spacetime]
{\includegraphics[width=59mm]{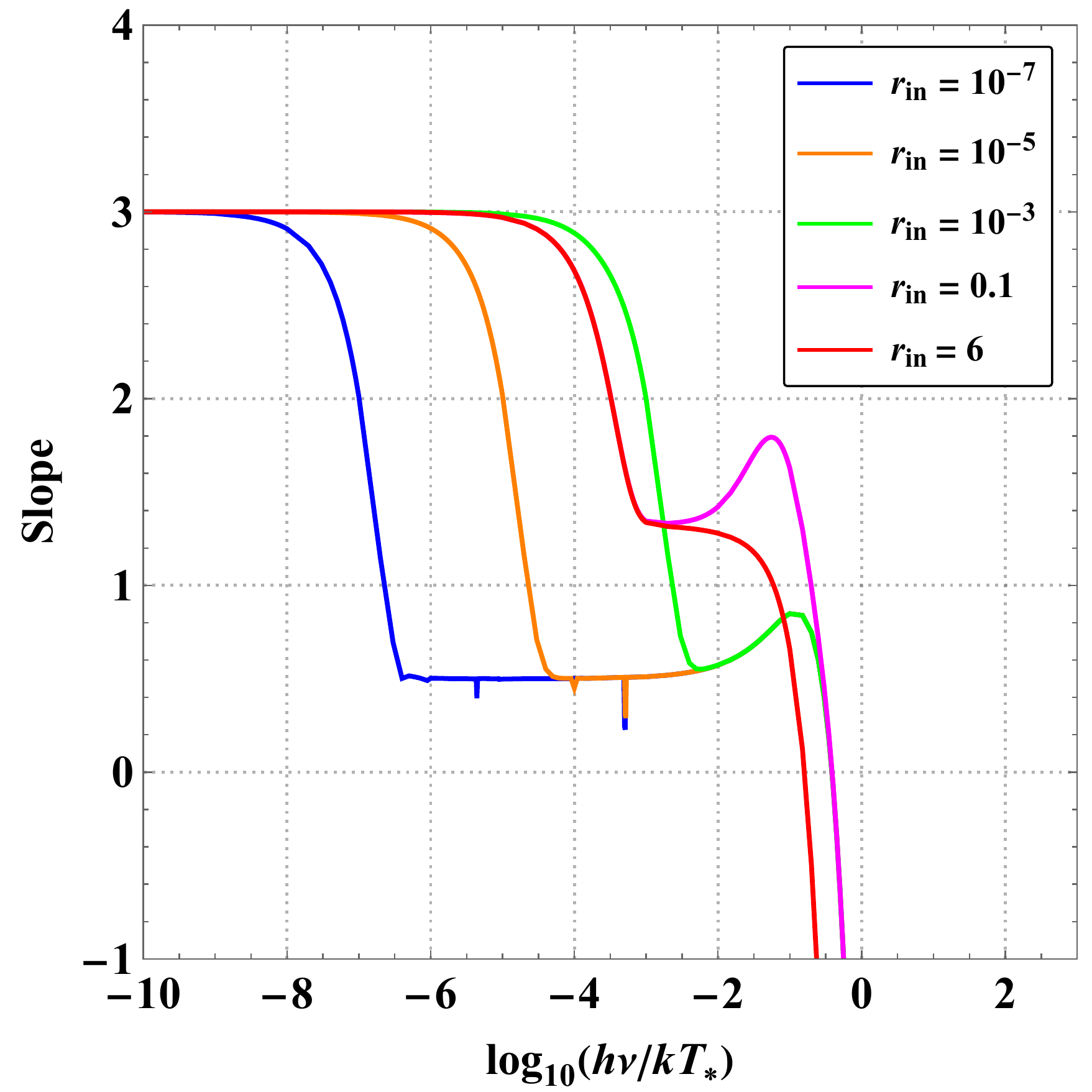}\label{fig:null_slope}}
\subfigure[JMN1 naked singularity ($M_0=0.3$, $r_b=6.6667$)]
{\includegraphics[width=59mm]{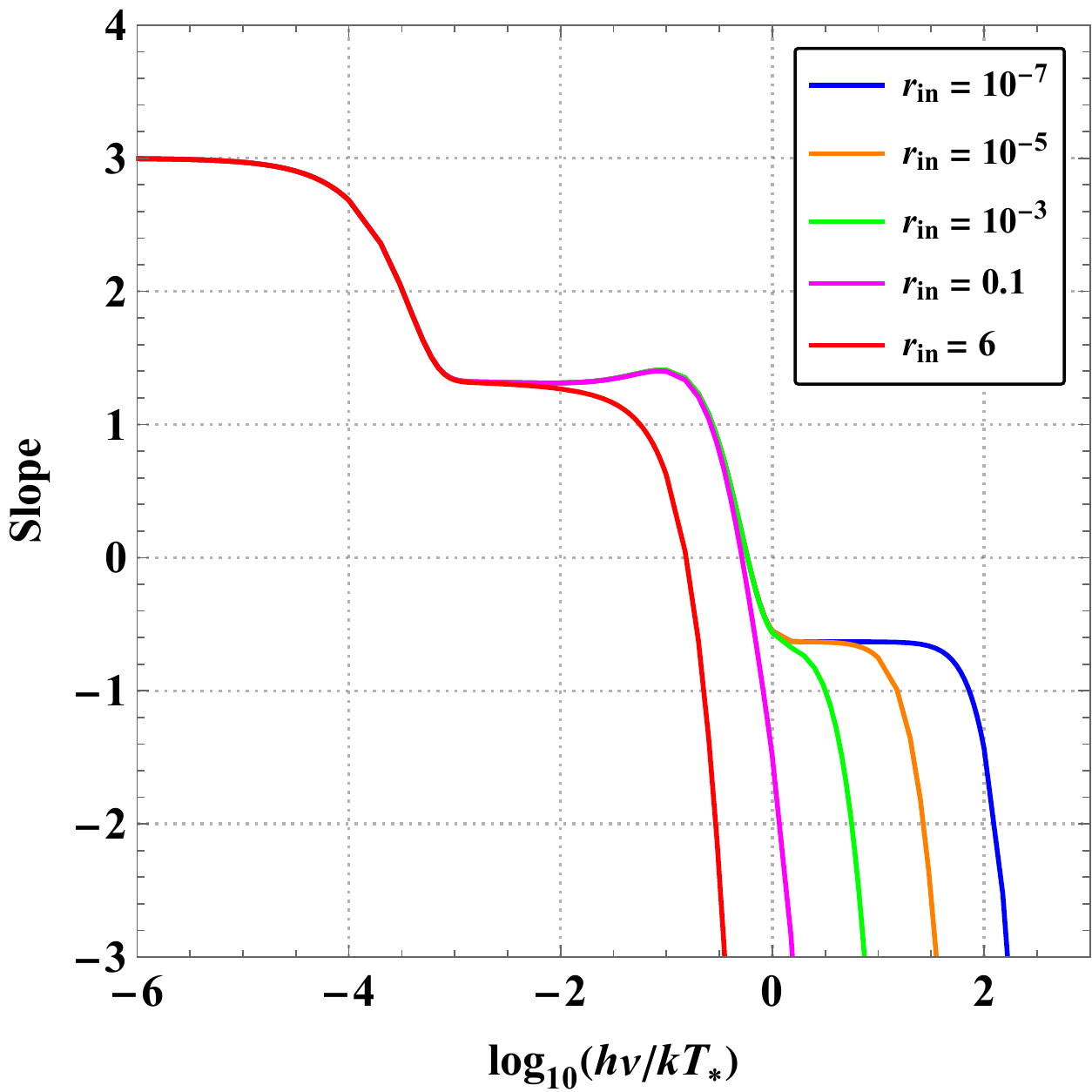}\label{fig:jmn1_slope}}
 \caption{The slopes of spectral luminosity distributions of radiation emitted from the accretion disks of Schwarzschild black hole, NNS, and JMN1 spacetimes. The slopes for NNS and JMN1 are shown for various values of $r_{in}$ as considered in Fig.~(\ref{fig:diff_rms_spectra}). The value of $r_{in}$ for the Schwarzschild black hole is $6 M_T$. }
\label{fig:L_slopes}
\end{figure*}

In figures Fig.~(\ref{fig:N_ratio}) and Fig.~(\ref{fig:J_ratio}), we present, respectively, the variation of the ratios of luminosities of NNS and JMN1 to luminosity of Schwarzschild black hole with frequency. The inner region of the accretion disk for NNS for $r_{in}=10^{-7}$ is approximately $10^{11}$ times more luminous than that of the Schwarzschild black hole of the same mass. The difference in the luminosity at low-frequency between NNS and Schwarzschild black holes decreases as $r_{in }$ is shifted to higher radii, representing that a large amount of low-frequency radiation is contributing from the inner regions of the disk. Around $h\nu/k T_{*}\approx1$, the luminosity in the Schwarzschild tends to zero, this is because the accretion disk ends at $r=6M_{T}$. However, for JMN1 the luminosity is very high around $h\nu/k T_{*}\approx1$, since its accretion disk extends up to the singularity. As a result, the ratio of luminosity ratios tends to infinity after $h\nu/k T_{*}\approx1$ (see Fig.~\ref{fig:J_ratio}).
%
\begin{figure*}[t]
\centering
    \subfigure[]
        {\includegraphics[width=65mm]{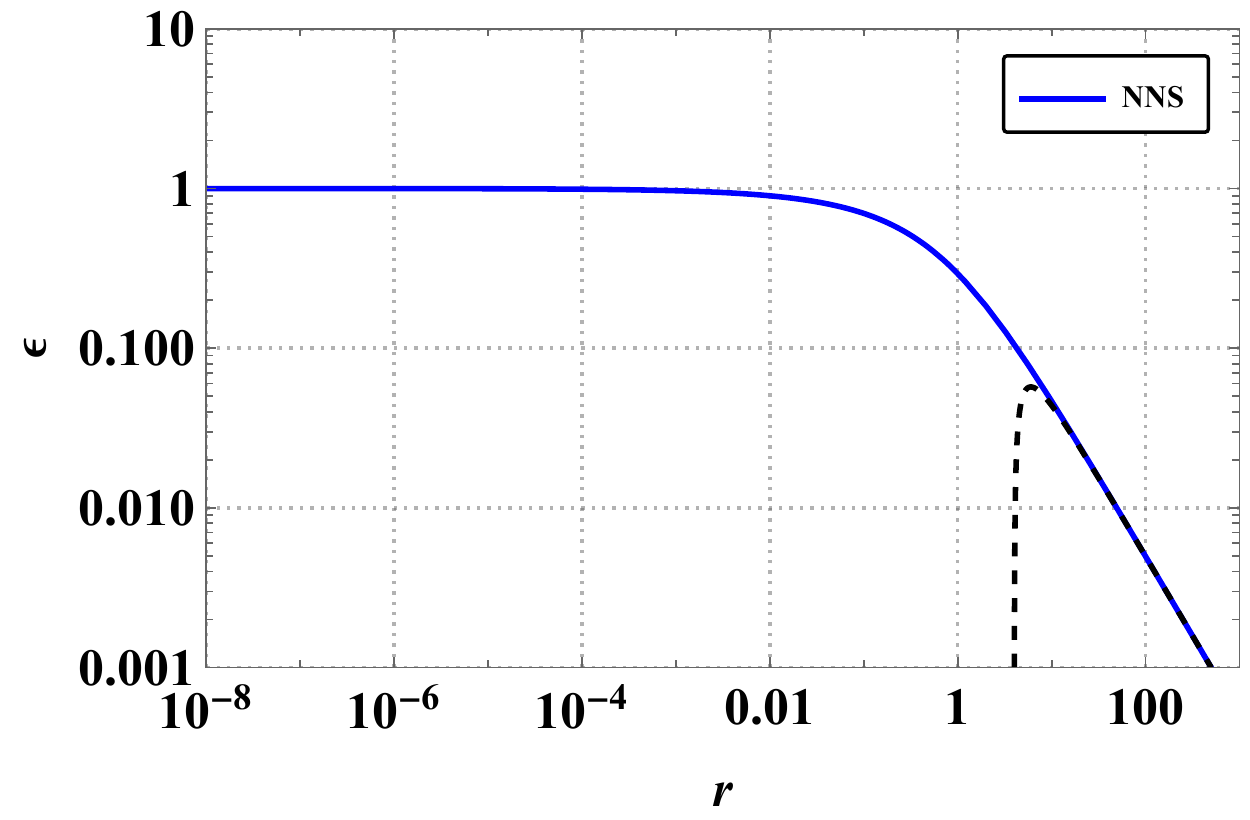}\label{fig:null_eff_slope}}
        \hspace{5mm}
    \subfigure[]
        {\includegraphics[width=65mm]{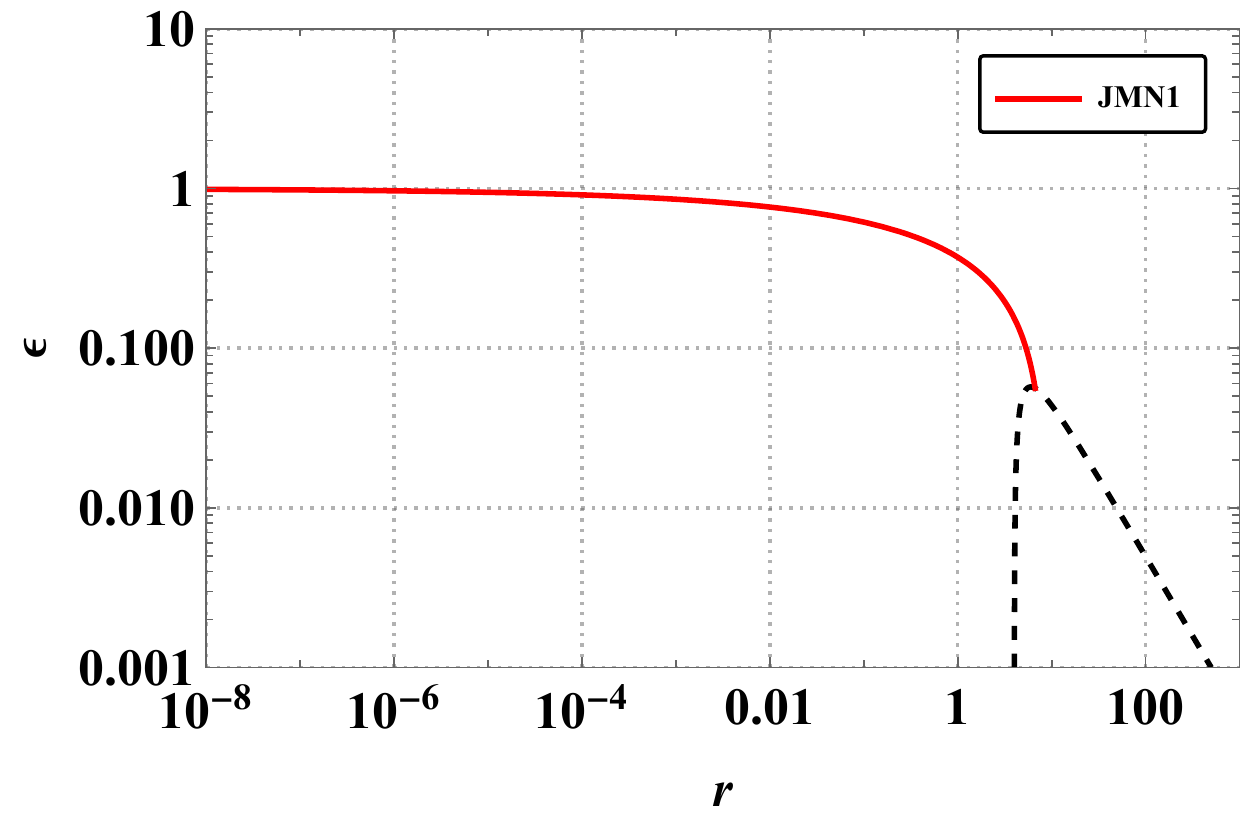}\label{fig:jmn1_eff}}
 \caption{The accretion efficiency as a function of radial distance $r$. The dashed black line corresponds to the Schwarzschild black hole. These plots show that the accreting matter loses energy as it approaches $r_{ms}$. For naked singularities, almost all of the rest mass is transformed into radiation, and hence the efficiency is nearly equal to $1$ at $r\to 0$. }
\label{fig:efficiency}
\end{figure*}
\subsubsection{Spectral luminosity distribution by introducing non-zero torque at the inner edge of the disk}
As discussed in (\ref{sec4}), Eq.~(\ref{flux}) can only be used to calculate the radiating flux if the ``zero-torque'' condition is followed at $r=r_{ms}$. For naked singularities where the accretion disc can reach up to the singularity, we cannot necessarily state that the torque acting on the accreting matter will totally drop to zero approaching the singularity. Due to the ultra-strong gravity in the regions near the singularity, the viscosity between the fluid layers in the accretion disc can be very high, which may induce torque in the particle. To examine the effect of non-zero torque at the inner edge of the accretion disk, we can modify Eq.~(\ref{flux}) by introducing a non-zero value of torque ($W^{r}_{\phi}$) at $r=r_{ms}$.  From Eqs.~(\ref{w1}) and (\ref{w2}), we see that the non-zero value of parameter $f$ should be considered at $r=r_{ms}$ to introduce the non-zero $W^{r}_{\phi}$. We denote the non-zero value of $f$ at $r_{ms}$ as $f_{ms}$. In that case, from Eq.~(\ref{f}), we get,
 \begin{equation}
        \frac{(E-\mathcal{W}  L)^2 }{ -\mathcal{W}_{,r}} f= \int_{r_{ms}}^{r} (E-\mathcal{W}  L)\,L_{,r} \, dr + f_{ms}\,.
    \end{equation}

  Therefore, the expression of flux radiated will be
 \begin{widetext}
    \begin{equation}
        \mathcal{F}(r)=-\frac{\dot{m}}{4\pi \sqrt{-g}}\frac{\mathcal{W}_{,r}} {(E-\mathcal{W} L)^2}\left(\int^r_{r_{\rm ms}}(E-\mathcal{W} L)L_{,r}dr + f_{ms}\right).
        \label{flux_fms} 
    \end{equation}
 \end{widetext}
Using Eq.~(\ref{flux_fms}), one can compute the energy flux of the electromagnetic radiation when a non-zero torque exists at the inner boundary of the disk. Fig.~(\ref{fig:diff_tq_spectra}) represents how the luminosity spectrum for NNS and JMN1 spacetimes varies with the parameter $f_{ms}$. From Figs.~(\ref{fig:null_diff_tq}), it can be noticed that a non-zero torque induces a rise in the spectral luminosity distribution of an NNS  at low frequencies. In the case of JMN1, however, we notice an increase in the spectral luminosity distribution at high frequencies in Fig.~(\ref{fig:jmn1_diff_tq}). Hence, the non-zero torque at the inner edge has a distinct influence on the luminosity spectra of NNS and JMN1 spacetimes. This distinguishing feature may aid in differentiating nulllike and timelike singularities considered here.
\subsubsection{Slopes of luminosity spectra of accretion disks}
Here, we examine the slopes of the spectral luminosity distribution of the radiations obtained in Fig.~(\ref{fig:diff_rms_spectra}) for Schwarzschild black hole and naked singularity models. The slope $J$ of the luminosity distribution spectra with respect to the frequency can be written as~\cite{Guo:2020tgv}
\begin{equation}
    J\equiv \frac{d[log_{10}(\nu\mathcal{L}_{\nu,\infty}/\dot{m})]}{d[log_{10}(h\nu/kT_*)]}\,.
\end{equation}\label{slope}

The numerical results of the slopes are shown in Fig.~(\ref{fig:L_slopes}) for $f_{ms}=0$ case. From Fig.~(\ref{fig:L_slopes}), it can be seen that at low frequency, the slope of spectral luminosity distribution is identical for all the models. Whereas at higher frequencies, there is a significant difference in the slopes for the models. We can use these features to distinguish the black hole and naked singularity models observationally. 
\begin{figure*}
\centering
\subfigure[Schwarzschild black hole]
{\includegraphics[width=65mm]{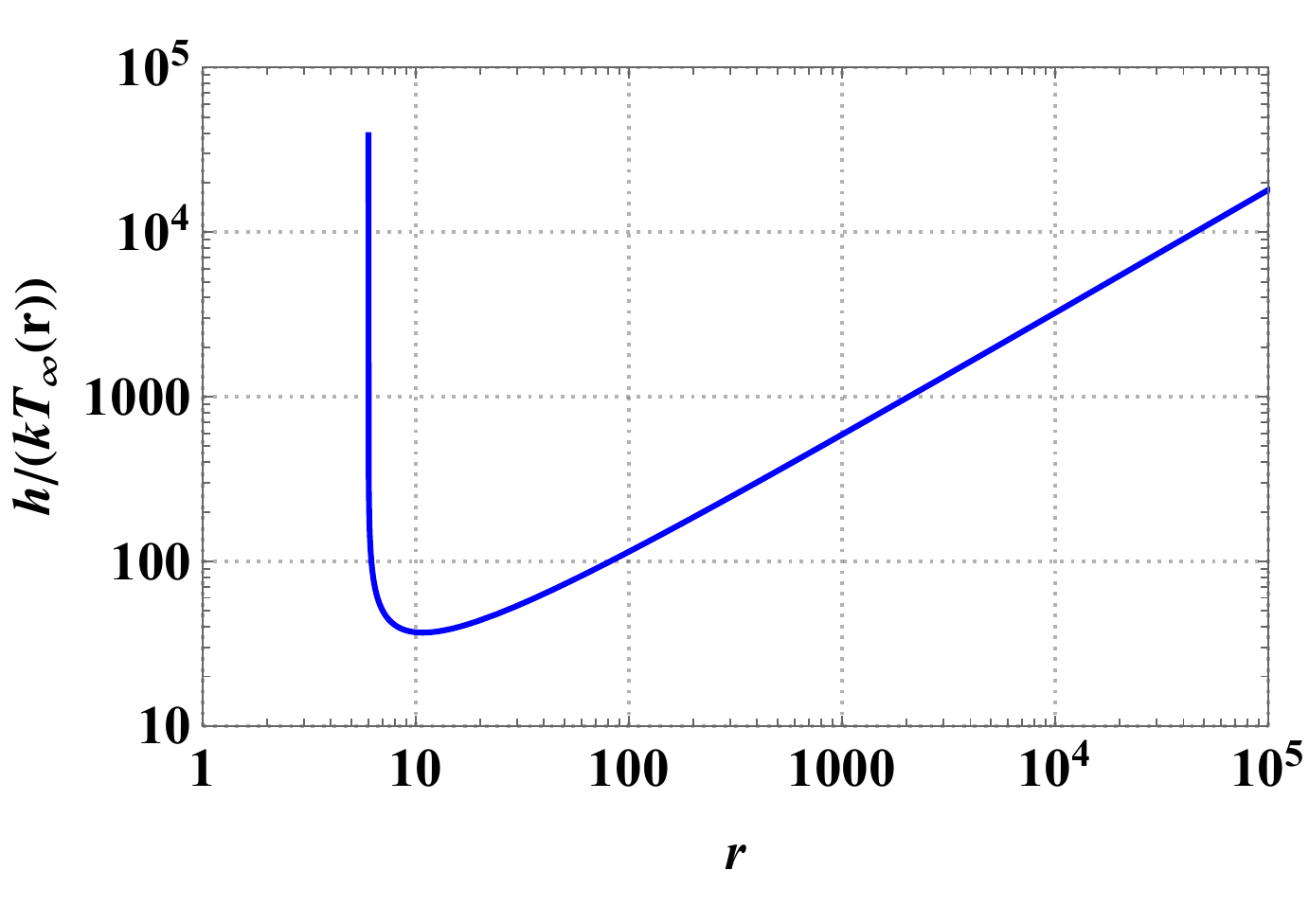}\label{fig:sch_temp}}
\hspace{5mm}
\subfigure[NNS spacetime]
{\includegraphics[width=65mm]{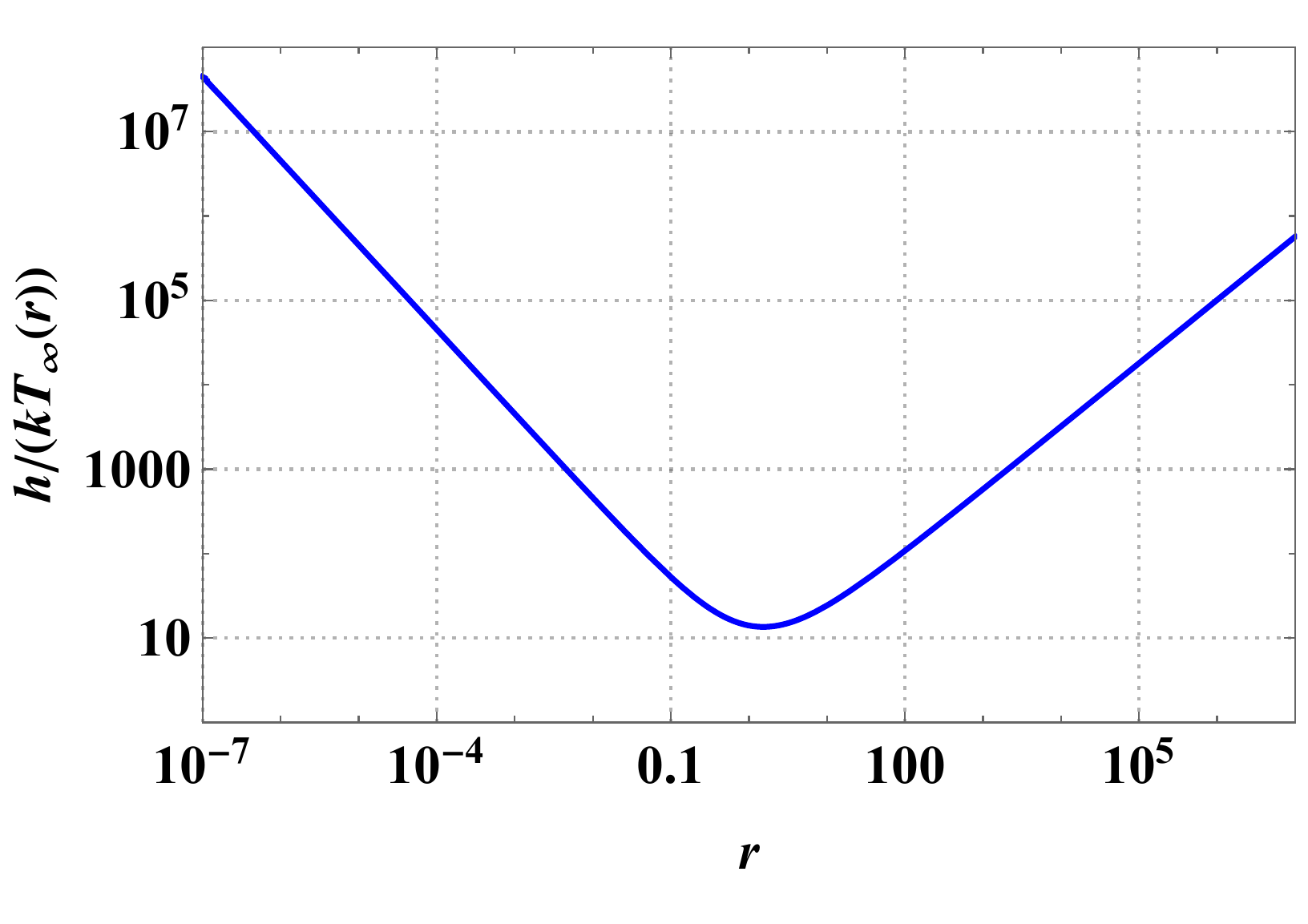}\label{fig:null_temp}}\\
\subfigure[JMN1 naked singularity ($M_0=0.3$, $r_b=6.667$)]
{\includegraphics[width=65mm]{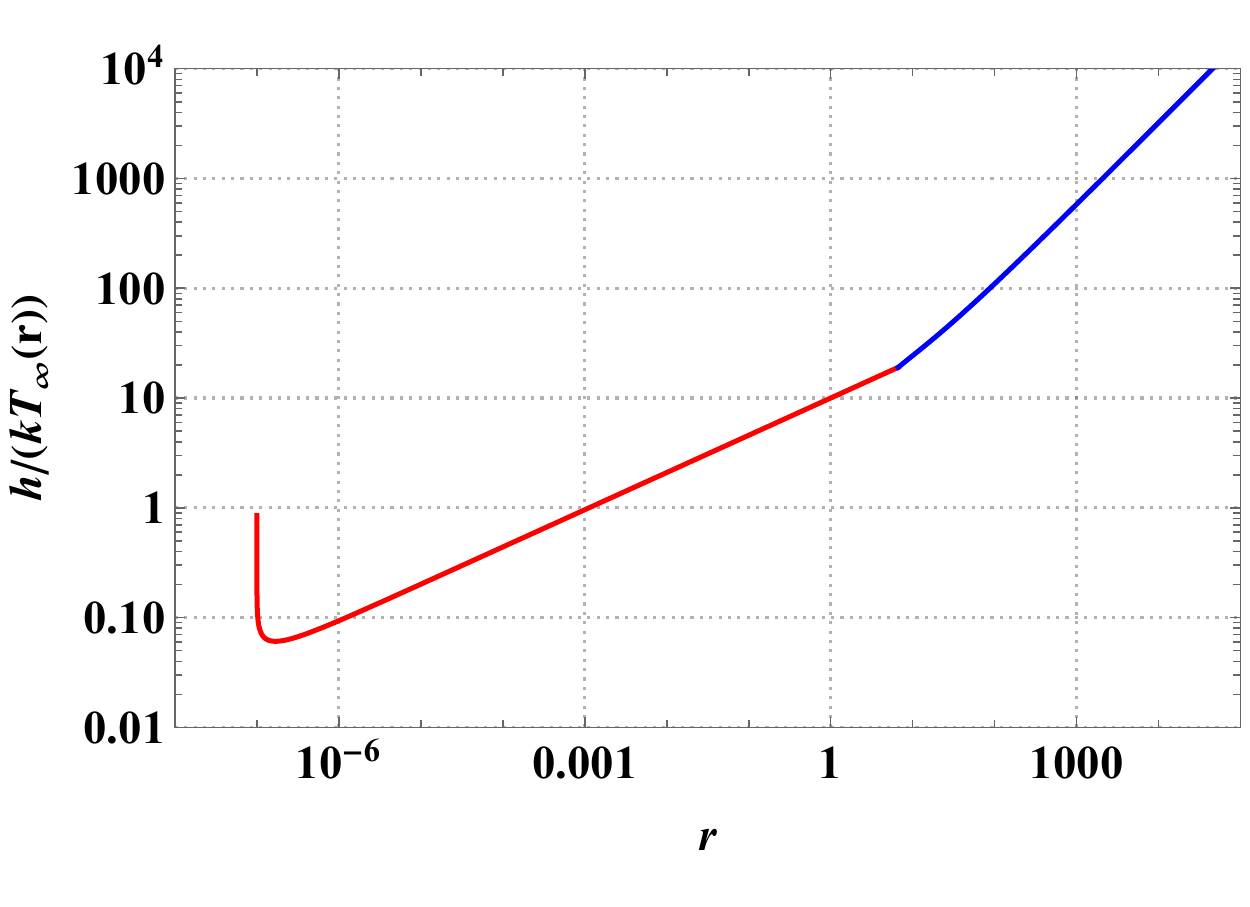}\label{fig:Jmn1_temp}}
 \caption{The plots of $h/(kT_\infty(r))$ vs. $r$ for Schwarzschild black hole, NNS and JMN1 models. The red and blue curves in Fig.~(\ref{fig:Jmn1_temp}) correspond to the interior and exterior spacetimes respectively. Here, $r_{in}$ is considered at $10^{-7}$ for NNS and JMN1 spacetimes.}
\label{fig:temp}
\end{figure*}
\subsection{Accretion efficiency}
 The variation of efficiency with radius $r$ is plotted in Fig.~(\ref{fig:efficiency}). The figure shows how the rest mass of accreting matter is converted into electromagnetic radiation as $r$ decreases from a large distance to the singularity. The efficiency of the accretion disk for Schwarzschild black holes is only $5.719\%$, since its accretion disk exists up to $r=6 M_T$. On the other hand, for NNS and JMN1, the specific energy $E$ goes to zero as $r\to0$, and hence their accretion efficiency is almost $100\%$. Thus, naked singularities are extremely effective engines for transforming mass-energy into radiation.

\subsection{Temperature profile}
As mentioned earlier, the high luminosity spectra are mainly contributed by the inner region of the accretion disk, which can be verified by the results plotted in Fig.~(\ref{fig:temp}). In Fig.~(\ref{fig:temp}), temperature distribution on the disk surface is shown for the same configurations that we investigated for the luminosity spectra in Fig.~(\ref{fig:spectra}). From Fig.~(\ref{fig:sch_temp}), it can be observed that, for the Schwarzschild black hole, the minimum of $h/(kT_\infty (r))$ is at $r\approx10.79$.  At $r=6$, $h/(kT_\infty (r))$ tends to infinity, because the inner edge of the disk is only up to $r=6M_T$. On the other hand, for the NNS  model, the minimum of $h/(kT_\infty (r))$ is at $r\approx1.53$, which is in the vicinity of the central singularity. The minimum value of $h/(kT_\infty(r))$ for the JMN1 naked singularity model is at $r \approx 1.7 \times 10^{-7}$, which lies very close to the singularity. These results imply that the radiation with the highest temperature is emitted by the regions in proximity to the inner edge of the disk. In addition, we can see that the maximum temperature is significantly higher for null and JMN1 naked singularities than for Schwarzschild black holes, indicating that the inner regions of accretion disks for naked singularities are hotter than those of black holes.
\section{Conclusion}\label{sec6}
In this paper, we study the properties of the thin accretion disks surrounding black holes and naked singularities using the Novikov-Thorne accretion disk model. We carry out an analysis of the thermal properties of the radiation emerging from the disk surface. As for the naked singularities, we considered null and JMN1 naked singularity spacetimes, for which no ISCO exists, and hence accretion disks can reach up to the singularity. We also analysed the physical properties of the particles in the disk, such as specific energy ($E$), specific angular momentum ($L$), and angular velocity ($\mathcal{W}$). The conclusions of the paper are summarised as:

\begin{itemize}
    \item We obtain the equation for calculating the emitting flux for the spacetimes joined at the junction through the matching hypersurface by employing the appropriate boundary conditions in the derivation of the flux equation. The comparative study of the energy flux emitted by the accretion disks of black holes and naked singularities reveals that the amount of flux emerging from the disk surface of naked singularities is higher compared to that of a black hole of the same mass and accretion rate. 
    \item From the electromagnetic spectra of the radiation, we find that, compared to the Schwarzschild solution, the spectral distribution from the inner regions of the disks in JMN1 spacetime is more luminous in the high-frequency range. In contrast, it is more luminous in the low-frequency range in null naked singularity spacetime. These unique imprints in the luminosity spectrum provide a tool for distinguishing spacelike, timelike, and nulllike singularities by using astrophysical measurements of the radiation spectra emerging from accretion disks.
    \item  In addition to that, we obtain the slopes of the luminosity distribution curve for black holes and naked singularity models. The results indicate a considerable difference in the slopes of black holes and naked singularities at high frequencies. Such distinct characteristics may, in principle, help to discriminate between naked singularities and black holes.
    \item In general, the ``zero-torque" condition is assumed at the inner edge of the accretion disk. We show how the luminosity spectra of naked singularities vary when a non-zero torque is introduced close to the singularity. The inclusion of the non-zero torque at the inner edge of the disk results in a distinguishable change in the luminosity profile of the null and JMN1 naked singularities. 
    \item Furthermore, we also compute the accretion efficiency $\epsilon$ with which the accreting matter is converted into radiation and find that the naked singularities are far more efficient than black holes in transforming mass-energy into electromagnetic radiation. While investigating the temperature distribution on the disk surface, we note that the accretion disks around naked singularities are much hotter than black holes. For null and JMN1 naked singularities, the maximum value of the temperature profile lies close to the singularity, indicating that the inner regions of the accretion disks are significantly hotter and more luminous than the outer ones.
\end{itemize}

The radiation emerging from the inner regions of accretion disks of compact objects is impacted by strong-gravity effects. Because of this, the emitting signal retains traces of the geometrical structure of the central compact object to a faraway observer. Therefore,  investigating the accretion processes of compact objects can reveal a lot about their physical properties.
  In this context, the typical features emerging in the luminosity spectrum of the accretion disks surrounding the naked singularities analysed in this work suggest the possibility of observing and differentiating naked singularities utilising astrophysical observations of the electromagnetic spectra from accretion disks.

\vspace{0.8cm}

\end{document}